\documentclass[11pt,twoside,a4paper]{article}

\usepackage[american]{babel}
\usepackage{bm}
\usepackage{amsfonts}
\usepackage{graphicx}
\usepackage{amsmath}
\usepackage{fancyhdr}
\usepackage{a4wide}

\usepackage{setspace}
\newcommand{\nn}{\nonumber}

\newcommand{\disp}{\displaystyle}

\usepackage{color}

\newcommand{\zd}{\delta}

\newcommand{\zs}{\sigma}

\newcommand{\ze}{\varepsilon}

\newcommand{\zo}{\omega}

\newcommand{\zy}{\psi}

\newcommand{\zh}{\eta}

\newcommand{\dif}{\; \textrm d }

\newcommand{\dsp}[2]{\frac{  \partial^2 #1 }{  \partial #2^2   } }
\newcommand{\dst}[2]{\frac{  \dif^2 #1 }{  \dif #2^2   } }
\newcommand{\dpp}[2]{\frac{  \partial #1 }{  \partial #2   } }
\newcommand{\dpt}[2]{\frac{  \dif #1 }{  \dif #2   } }

\newtheorem{teor}{Theorem}
\newtheorem{lemma}{Lemma}[section]

\begin{document}

\begin{center}
\Large{\textbf{Quantum motion with trajectories: beyond the Gaussian beam approximation.}}\\
	\small{O. Morandi}	\\
\vskip0.5cm
	\textit{\textsf University of Florence  \\
Firenze, Italy}	
\vskip0.5cm
	\textit{omar.morandi@unifi.it}
\end{center}

\begin{center}
\begin{minipage}[h]{0.8\textwidth}
\section*{\textsf{Abstract}}
 \small
\textsf{A quantum model based on a Euler-Lagrange variational approach is proposed. In analogy with the classical transport, our approach maintain the description of the particle motion in terms of trajectories in a configuration space. Our method is designed to describe correction to the motion of nearly localized particles due to quantum phenomena. We focus on the simulation of the motion of light nuclei in ab initio calculations. Similarly to the Gaussian beam method, our approach is based on a ansatz for the particle wave function. We discuss the completeness of our ansatz and the connection of our results with the Bohm trajectories approach.}

\end{minipage}
\end{center}
\vspace{1cm}
\normalsize

\section{Introduction}

During the last decade, several numerical methods based on ab initio approaches have been developed. The improvement of the theoretical models and the numerical algorithms has eased the systematic investigation of the dynamics of the molecules and of the electronic and optical properties of the nanostructures \cite{Dammak_09,Garashchuk_13,Basile_13,Muscato_16,Jacoboni_14,Coco_16,Camiola_14,Sellier_15,Sprengel_17}.  

In  the framework of the ab initio quantum mechanical methods, a system containing $N$ protons and $N'$ electrons is described by a single many-body wave function $\Psi (\mathbf{R},\mathbf{r},t)$. We denote by $\mathbf{r}$ ($\mathbf{R}$) the collective variable containing the coordinates of all the electrons (nuclei). Even for a small number of particles the wave function $\Psi $ is extremely hard to compute. This is especially true if the dynamical evolution of $\Psi $ is addressed. In order to reduce the complexity of the problem, various approximations of $\Psi (\mathbf{R},\mathbf{r},t)$ have been proposed in order to disentangle to some extent the electronic and the nuclear component of the wave function \cite{Basile_13,Abedi_10}.  
According to  Born-Oppenheimer ansatz, the total wave function of electrons and nuclei is factorized in terms of  the product of the electronic wave functions ($\Phi_i$) and the atomic wave functions ($\Theta_i$)
\begin{align*}
\Psi(\mathbf{R},\mathbf{r},t) =\sum_{i} \Phi_i(\mathbf{R},\mathbf{r},t)\Theta_i (\mathbf{R},t)\; .
\end{align*}
A very popular approximation proposed firstly by Hartree, is to replace  $\Psi(\mathbf{R},\mathbf{r},t)$ by a single term $\Psi(\mathbf{R},\mathbf{r},t)\simeq \Phi (\mathbf{R},\mathbf{r},t)\Theta  (\mathbf{R},t)  $. 
The Hartree approximation is implicitly assumed by the majority of the ab inito methods based on the Kohn-Sham density functional theory (DFT) \cite{Curchod_13}. 
The second and major approximation is to describe only the electronic degrees of freedom by a quantum model. The wave function $\Phi (\mathbf{R},\mathbf{r},t)$ is calculated by the solving the quantum Kohn-Sham equation. The nuclei are considered point-like particles which move along the classical trajectories $\mathbf{R}(t)$ obtained by solving the Newton equation. Such a two-scale modeling, i. e. classical treatment for the nuclei and quantum description for the electrons, is motivated by the fact that the De Broglie length associated to a proton is usually very small compared with  De Broglie length associated to one electron under similar conditions of temperature and pressure.

Recent theoretical and experimental studies show the emergence of physical conditions for which such an approximation is violated. They suggest that the behavior as a quantum particle of slightly bound protons or of the hydrogen is at the origin of some complex phenomena. 
As an example, various chemical reactions are understood in terms of electronic transfer processes assisted by non adiabatic quantum transitions of protons. Such mechanisms may explain some biological processes related to the photosynthesis and to the DNA biosynthesis  \cite{Migliore_14,Hudock_07}.
Studying the phase transition of the ice of water by DFT models, Bronstein et al. have shown that in order to obtain the experimental value of the transition pressure, it is essential to include in the numerical model the quantum mechanical delocalization of the protons \cite{Bronstein_14}.

Motivated by these results,  the study of the mathematical models for the quantum transport of protons at the nanometric scale has seen renewed interest.
The development of new ab initio methods for the simulation of complex systems that include the description of the quantum motion of light atoms, has a wide array of potential applications. 
In a large number of molecular configurations, the hydrogen atoms are trapped in local minima of the electrostatic potential whose barriers have thickness of few nanometers.
%
In such cases, the zero point energy of the protons and the quantum tunneling play a relevant role to determine the equilibrium configuration of the system.

New concepts such as the deterministic and the stochastic quantum trajectories have been proposed in order to develop new ab initio methods that go beyond the classical description of the atomic nuclei \cite{Maddox_01,Wang_15}. One of the major advantages provided by a method that extends the concept of classical trajectory to the quantum mechanical context, is that it could be integrated through minimal changes in the solvers already developed for the molecular dynamics. 
The Bohm interpretation of the quantum mechanical nature the particles extends in a rigorous way the concept of the deterministic trajectory of a particle. The Bohm approach  is potentially able to deal with any quantum phenomena. Nowadays, methods based on the Bohm formalism  are very popular. However, the application of the Bohm theory to a real situation suffers from serious limitations due to the fact that the calculation of the quantum Bohm trajectories is extremely challenging. 
The first attempts in this direction, have been proposed by Tully \cite{Tully_90} and Wyatt \cite{Wyatt_05}. The dynamics of the nuclei is described by a single or by a bundle of quasi-classical trajectories that are evaluated selfconsistently with the electronic density. 
Such methods have stimulated the study of few general methods such as the multiconfiguration time-dependent Hartree method \cite{Beck_00}, the conditional Born-Oppenheimer dynamics \cite{Albareda_15} and the Bohm trajectories extended to the complex plane \cite{Goldfarb_06}. Moreover, also stochastic methods have been addressed. Based on the classical concept of Brownian and Langevin dynamics, such stochastic methods are able to reproduce the quantum statistical properties of protons in harmonic traps \cite{Ceriotti_09}.  

In this paper, we propose a quantum model based on a Euler-Lagrange variational approach.  Our method is designed to describe the motion of nearly localized particles. Similarly to the Gaussian beam methods \cite{Jin_14}, we assume that the particle wave function is well described by a single Gaussian modulated by a polynomial. Our ansatz for the particle wave function contains a complete set of parameters. We derive the evolution equation of such parameters by considering an integral formulation of the evolution equations. Our method show a strong analogy with the Bohm approach. The integral form of the equations suggests a strategy which is able to tackle with the divergences of the Bohm potential. 

\section{Euler-Lagrange variational approach}
We derive the motion of a quantum particle by using a variational approach.  We assume that the particle is described by a generalized Gaussian wave packet parametrized by a set of time dependent numbers $\xi_n$. We formulate the particle evolution in terms of a Euler-Lagrange problem. By minimizing the action of the quantum Lagrangian of the particle, we derive the evolution equations for the parameters $\xi_n$. In the following we show that our ansatz is completely general so that every L$^2$ function can be expressed by the $\xi_n-$parametrized Gaussian wave packet. However, we remark that our main concern is to derive approximated models that describe the evolution of  quasi-classical  particles \cite{mor_JPA_10,Mor_12JMP}. 
In standard conditions, a heavy quantum particle has small De Broglie wave length and is characterized by few parameters namely the mean position, velocity and dispersion.
Our method is designed to describe the motion of well localized wave packets with Gaussian dispersion and we focus on neutron and proton motion.
As a first approximation, the wave function of a quasi-classical particle is usually approximated by a minimum uncertainty Gaussian profile 
\begin{align}
\zy (x,t) =&\sqrt[4]{\frac{\zs(t)}{\pi}} e^{ - \frac{ [x-s(t)]^2\left[\zs(t)- i \zs_i(t) \right] }{2}}e^{ i \left[ \phi_0(t)+  p(t)   (x-s(t))    \right]   } \;.\label{ansatz exa}
\end{align}
Here, in order to introduce our model, we have parametrized the particle wave function $\psi$ by inserting few real numbers which are associated to the main physical quantities of the particle. In particular,  $s$ is the mean position, $p$ the mean momentum, $\phi_0$ the phase and $\zs$, $\zs_i$ are, respectively, the real and imaginary part of the inverse of the standard deviation. By assuming that the ansatz of Eq. \eqref{ansatz exa} is valid for a sufficiently long time interval, by evaluating the evolution of $s$, $p$, $\phi_0$, $\zs$ and $\zs_i$, it is possible to obtain an approximate evolution of the particle wave function $\psi$. In the following, we generalize the  ansatz given in Eq. \eqref{ansatz exa} by defining a more general set of parameters. 
 
We consider the quantum mechanical evolution of a particle in the presence of the potential $U(x)$. 
\begin{align}
i \dpp{\zy}{t} =& \left(- \frac{1}{2}\dsp{}{x} +U(x) \right) \zy\;. \label{sch ini}
\end{align}
Hereafter, the Planck constant $\hbar$ and the particle mass are set to one. For the sake of simplicity, we have considered the one-dimensional motion. 
According to the Euler-Lagrange formalism, we define a suitable quantum Lagrangian $L$ and we formulate the quantum motion in terms of a variational problem for $L$ \cite{Haas_09}.  We define the quantum Lagrangian
\begin{align*}
L= & \int_\mathbb{R}\mathcal{L} (\zy,\zy^\dag,\partial_x\zy, \partial_x\zy^\dag, \partial_t\zy, \partial_t\zy^\dag , x,t) \dif x\;,
\end{align*} 
where the Lagrangian density is
\begin{align*}
\mathcal{L}  =&  -\frac{i}{2} \left(\partial_t \zy^\dag \zy -  \zy^\dag \partial_t\zy\right) - \frac{1}{2} \partial_x \zy^\dag \partial_x \zy-\; U  \zy^\dag  \zy  \;.
\end{align*}
Equation \eqref{sch ini} is formally equivalent to the Euler-Lagrange equation  
\begin{align}
\frac{\partial \mathcal{L}}{\partial \zy^\dag } =&\partial_t \frac{\partial \mathcal{L}}{\partial \left(\partial_t \zy^\dag\right) } +
\partial_x \frac{\partial \mathcal{L}}{\partial \left(\partial_x \zy^\dag\right) }\label{Eu-Lag} 
\end{align}
where, according to the standard field quantization procedure, $ \zy$ and $\zy^\dag$ are considered two independent fields.  We write the wave function $\zy$ as follows
\begin{align}
\zy (x) =&\sqrt{P(x-s(t),t)}  e^{ - \frac{ (x-s(t))^2}{2}  \zs(t) +i  \chi  (x-s(t),t)  } \;. \label{ansatz sch1}
\end{align}
Here,  $P$ and $\chi$ are polynomials which depend on time $P=P(x,t)$, $\chi=\chi(x,t)$ and $s=s(t)$, $\zs=\zs(t)$ are two parameters.
We treat $\zy^\dag$ as the complex conjugate of $\zy$ \cite{Haas_09}.
The quantum Lagrangian becomes
\begin{align}
L =&\int_\mathbb{R}   \left[ \textrm{Im} \left(  {\zy}^\dag  \partial_t\zy\right) - \frac{1}{2} |\partial_x \zy|^2- U  |  \zy|^2 \right] \dif x \nn \\
%
=&\int_\mathbb{R} \left[  P   \dpt{\chi}{x} \dot{s} - P   \dot{\chi} 
- \frac{ \zs^2 x^2P}{2} -  \frac{1}{8P} \left( \dpt{P}{x} \right)^2  + \frac{ x\zs }{2}\dpt{ P}{x}  - \frac{ P  }{2}  \left( \dpt{\chi}{x} \right)^2 -U_{[s]}   P   \right]   e^{- x^2 \zs  }  \dif x\;. \label{lagr_gen}
\end{align}
Here, $U_{[s]}$ denotes $U(x-s)$, Im the imaginary part and the dot the time derivative. 
\subsection{Example: Gaussian wave}
We illustrate the application of the Euler-Lagrange method by a simple example. We consider the standard Gaussian wave ansatz expressed by Eq. \eqref{ansatz exa} which can obtain as a particlar case of Eq. \eqref{ansatz sch1} for $P=  a_0 \zs^{1/2} \pi^{-1/2}$ and $\chi= \phi_0+  p  x+      \frac{\zs_i}{2}x^2  $. We write Eq. \eqref{lagr_gen} in terms of the Lagrangian parameters $\xi\equiv(a_0,s,\phi_0,p,\zs_i,\zs)$
\begin{align*}
L =&    a_0     \left[   - \dot{\phi}_0    +  \dot{s} p     -   \frac{ p^2 }{2}    -\frac{ \dot{\zs}_i   +  \zs^{2}_i +\zs^2 }{4 \zs}     -\sqrt{\frac{\zs }{\pi } } \int_\mathbb{R}U (s-x)   e^{- x^2 \zs  }  \dif x  \right]  \;.
\end{align*}
The Euler-Lagrange equations are obtained by evaluating the stationary quantum action 
$
\delta \int L \dif t=0
$
\begin{align}
 \dpt{}{t} \frac{\partial L}{\partial \dot{\xi}_n } =& \frac{\partial L}{\partial \xi_n }\;. \label{Eu-Lag2} 
\end{align}
We obtain the following closed system of equations
\begin{align}
\dot{a_0} =&0 \\
\dot{s} =&    p \\
  \dot{p} =& \sqrt{\frac{\zs }{\pi } } \int_\mathbb{R}\dpt{ U(s-x)}{x}    e^{- x^2 \zs  }  \dif x  \label{ex p dot}\\
\dot{\zs} =& - 2  \zs_i \zs  \label{ex zs dot}\\
\dot{\zs}_i=  &         \zs^{2} -   \zs^{2}_i  - 2 \sqrt{\frac{\zs^3 }{\pi } } \int_\mathbb{R}\left( 2     \zs x^2  - 1  \right)    U (s-x)e^{- x^2 \zs  }  \dif x\label{ex zsi dot}\\
  \dot{\phi}_0  =&        \frac{ p^2}{2}    -\frac{\left(\zs -\zs_i\right)^2}{4\zs}       - \sqrt{\frac{\zs }{\pi } }\int_\mathbb{R}    U (s-x)e^{- x^2 \zs  }  \dif x  \;.\label{ex phi dot}
\end{align}
The first equation ensues that the L$^2$ norm of the wave function is conserved. In fact, from Eq. \eqref{ansatz sch1} we have  $\left\| \zy \right\|_2 = a_0^{1/2} $. 
It is interesting to consider the  harmonic potential $U=  \frac{\zo^2}{2} x^2$. Equations \eqref{ex p dot}-\eqref{ex phi dot} simplify 
\begin{align}
   \dot{p} =& 
       \zo^2 s  \\
 \dot{\zs}_i=  &         \zs^{2} -   \zs^{2}_i  -\zo^2 \label{ex zsi dot gaus}\\
  \dot{\phi}_0  =&     \frac{ p^2}{2}   - \frac{\zo^2 s^2}{4} -\frac{\zs}{2}\;.
 \end{align}
We used
\begin{align*}
\zs^{3/2}\pi^{-1/2} \int_\mathbb{R}\left(   2      \zs x^2  - 1  \right)    U (s-x)e^{- x^2 \zs  }  \dif x   =  &  \frac{\zo^2}{2}\;.
\end{align*}
\subsection{General case}
We discuss now the Euler-Lagrange problem in a more general context. We expand the wave function $\zy$ on a complete set of functions. We write $P$ and $\chi$ as follows  
\begin{align}
P (x,t)=&\frac{1}{\pi^{1/4} }\sum_{n=0}^\infty h_n^\zs (x) a_n(t) \label{expans P}\\
\chi (x,t)=&\pi^{1/4}\sum_{n=0}^\infty h_n^\zs (x) \chi_n(t)\;. \label{expans chi}
\end{align}
Here, $a_n$ and $\chi_n$ are the coefficients that parametrize the solution and the symbol $ h_n^\zs (x)$  denotes the normalized Hermite polynomials 
\begin{align}
h_n^\zs (x)&\equiv \frac{\zs^{1/4}}{\sqrt{2^n n!\sqrt{\pi}}}  H_n\left( x\sqrt{\zs} \right)  =(-1)^n  \frac{\zs^{1/4-\frac{n}{2}}}{\sqrt{2^n n!\sqrt{\pi}}} e^{x^2\zs} \frac{\dif^{(n)}}{\dif x^{n}} e^{-x^2\zs }\;.\label{def hn-zs}  
\end{align}
For the sake of clarity, in the previous expression  we have expressed $h_n^\zs (x)$ in terms of the  more standard definition of Hermite polynomials
\begin{align}
H_n (x)& \equiv (-1)^n e^{x^2 } \frac{\dif^{(n)}}{\dif x^{n}} e^{-x^2 }\;. \label{H hermite}
\end{align}
It is well known that the polynomials $h_n^\zs$ form a complete basis set for the inner product
\begin{align}
\langle  h_n^\zs, h_m^\zs \rangle_\zs \equiv \int_\mathbb{R}h_n^\zs (x) h_m^\zs (x) e^{-x^2 \zs} \dif x = & \zd_{n,m}\;. \label{ortho h}
\end{align}
The symbol $\zd$ denotes the Kronecker delta. 
For future references, we state here some properties of the Hermite polynomials
\begin{align}
\dpt{h_n^\zs}{x} =& \sqrt{2 n \zs}\; h_{n-1}^\zs \label{der hn-zs}\\
\sqrt{\zs} x\; h_n^\zs =&  \sqrt{\frac{n+1}{2}   }\; h_{n+1}^\zs+\sqrt{\frac{n}{2}   }\; h_{n-1}^\zs\label{svil x hn-zs}\\
\sqrt{2n \zs }\; h_{n}^\zs  =&\left( 2 \zs x - \dpt{}{x}\right)  h_{n-1}^\zs    \label{hn-zs3} \\
\dpt{h_n^\zs}{\zs} =&      \frac{2n+1}{4\zs}  h_{n}^\zs+\frac{\sqrt{n (n-1) }}{2\zs} \;  h_{n-2}^\zs  \label{der zs hn-zs}\;.
%
\end{align}
The following scaling property will be very useful
\begin{align}
h_n^\zs (x)  =  h_n^1 \left( x\sqrt{\zs} \right) \zs^{1/4} \; . \label{scal h zs}
\end{align}

\subsection{Completeness of the expansion }\label{sec comp exp}

By using the expansion of Eqs. \eqref{expans P}-\eqref{expans chi} we have parametrized the particle wave function by the set $\xi\equiv\{s,p,\zs_i,\zs,a_n,\chi_n\}$. 
The Euler-Lagrange equations \eqref{Eu-Lag2} for such  coefficients are obtained by straightforward computation of the Lagrangian \eqref{lagr_gen}. In particular, by analysing the number of coefficients that we have introduced, it is easy to recognize our choice of  coefficients is redundant. In fact, since the Hermite polynimials form a complete set, it is clear that every function $P$ and $\chi$ can be expanded according to, respectively Eq. \eqref{expans P}  and Eq. \eqref{expans chi}. The functions $P$ and $\chi$ are sufficient to determine the modulus and the phase of $\zy$. However, our ansatz contains two additional parameters $\zs$ and $s$. This suggests that we can fix the values of two parameters among all the others. Direct computation shows that the quantum Lagrange equations are considerably simplified in a particular case, namely if the coefficients $a_1$ and $a_2 $ of the expansion \eqref{expans P}, are identically zero.  In the following  we will show that every L$^2$ function can be expanded according to Eqs. \eqref{ansatz sch1},\eqref{expans P},\eqref{expans chi} under the constraint $a_1=0$ and $a_2 =k$ where $k$ is given.

By writing $\psi$ in polar coordinates $\psi=|\psi|e^{i\chi}$, it is clear that  constraints on $a_1$ ad $a_2 $ concern only the expansion of the modulus $|\psi|$. In order to show the completeness of our expansion, it is sufficient to show that the modulus   $|\psi(x)|^2=P(x-s) e^{ - (x-s)^2  \zs} $ can expressed  by 
\begin{align}
\left|\zy (x)\right|^2 =& \frac{e^{ - (x-s)^2 \zs}}{\pi^{1/4}}   \left( h_0^\zs a_0+ h_2^\zs k +\sum_{n=3}^\infty h_n^\zs (x-s) a_n  \right) \;.\label{exp psiquad}
\end{align}
It is well known that for every choice of the parameters  $s\in \mathbb{R}$ and $\zs\in \mathbb{R}^+$ the set $ \{h^\zs_n(x-s)e^{-(x-s)\zs/2} |\; : n\in \mathbb{N} \}$, is complete in the  space L$^2(\mathbb{R})$. Equivalently,   the set $ \{h^\zs_n(x-s)  |\; : n\in \mathbb{N} \}$  is complete in the space of the polynomials. 
The expansion \eqref{exp psiquad} is complete if and only if for every polynomial  $p$, we can find  $s\in \mathbb{R}$ and $\zs\in \mathbb{R}^+$ such that the second and third coefficient of the Hermite expansion of $p$  are prescribed. This is ensured by the following
\begin{lemma}
	Let $p$ be a non-negative polynomial of degree $2n$ with $n\in\mathbb{N}$. There exist $k_M\in \mathbb{R}$ such that for every $k > k_M$ it is possible to find $\zs\in \mathbb{R}^+$ and $s\in \mathbb{R}$ such that 
	\begin{align}
 &\langle h_1^\zs(x-s), p \rangle_\zs =0  \label{p1_lemma}\\
 &\langle h_2^\zs(x-s), p \rangle_\zs =k \;.\label{p2_lemma}
	\end{align}
\end{lemma}
\emph{Proof.}
%
%
We prove the Lemma by fixed point argument.  We define 
\begin{align}
p_1(s,\zs) \equiv& \frac{\zs^{-1/4} \pi^{-1/4}}{\sqrt{2}} \langle h_1^\zs(x-s), p \rangle = \frac{\zs^{3/2}}{\pi^{1/2}}  \int_{-\infty}^\infty (x-s) p (x)  e^{-(x-s)^2\zs} \dif x \;, \label{p1_lemma-proof}\\
p_2(s,\zs) \equiv&\zs^{5/4} \sqrt{2}\pi^{1/4} \langle h_2^\zs(x-s), p \rangle=\zs^{3/2}  \int_{-\infty}^\infty \left[  2\zs (x-s)^2- 1    \right]p (x)  e^{-(x-s)^2\zs} \dif x   \;,\label{p2_lemma-proof}
\end{align}
We indicate by $s=G_1(\zs)$ the solution of $p_1(s,\zs) =0$  (the variable $s$ is expressed as a function of  $\zs$) and $\zs=G_2( s)$ the solution of  $p_2(s,\zs) =0$  (the variable $\zs$ is expressed as a function of $s$). We prove that the following map is well defined
\begin{align*}
\left\{\begin{array}{llr}
s_i=&G_1(\zs_{i-1})\\
\zs_i=&G_2(s_i) & 
\end{array}\right. i=1,2,\ldots
\end{align*}
for every initial value $\zs_0$. We show that the images of the maps $G_1$ and $G_2$ are compact. This ensures the existence of a convergent subsequence of the set $(s_i,\zs_i)$ and concludes the proof of the Lemma. 

We start by showing that if we fix $\overline\zs \in \mathbb{R}^+$, $p_1(s,\zs) =0$ admits always a solution $s=G_1(\overline{\zs})$. 
If $p(x)$ is constant, the solution is $s=0$. We assume that $p(x)$ is not  constant. From Eq. \eqref{p1_lemma-proof}, $p_1(s,\zs) =0$ can be written as
\begin{align}
\dpt{}{s}K_{\overline{\zs}}(s) =0\;,\label{dds a1eq0}
\end{align}
where $K_{\overline{\zs}}(s) \in C^\infty (\mathbb{R})$ is given by
\begin{align}
 K_{\overline{\zs}}(s)\equiv \int_{-\infty}^\infty p (x)  e^{-(x-s)^2\overline{\zs}} \dif x \;. \label{dds a1eq0}
\end{align}
It is easy to verify that for every positive defined polynomial of degree $2n$ there exist $\overline{x}$ and $M>0$ such that, for every  $x \in \mathbb{R}$, $|x|\geq \overline{x} \Rightarrow
p (x) \geq M x^{2n}$. For $s>\overline{x}+1 $ we have 
\begin{align*}
K_{\overline{\zs}}(s)= & \int_{|x|>\overline{x}} p (x)  e^{-(x-s)^2\overline{\zs}} \dif x  + \int_{|x|< \overline{x}} p (x)  e^{-(x-s)^2\overline{\zs}} \dif x    
\geq   M\int_{|x|>\overline{x}}   x^{2n} e^{-(x-s)^2\overline{\zs}} \dif x  +   \sqrt{\frac{\pi}{\overline{\zs}}}\; \min_{|x|< \overline{x}}  p (x)  \\ 
\geq & Me^{-\overline{\zs}} \int_{s-1}^{s+1}   x^{2n}  \dif x+ \sqrt{\frac{\pi}{\overline{\zs}}}\; \min_{|x|< \overline{x}}  p (x)   
= \frac{Me^{-\overline{\zs}}}{2n+1} \left[\left( s+1 \right)^{2n+1}-\left( s-1 \right)^{2n+1}\right] +  \sqrt{\frac{\pi}{\overline{\zs}}}\; \min_{|x|< \overline{x}}  p (x)  \;. 
\end{align*}
We used $e^{-x^2\zs}\geq e^{-\zs}\chi_{|x|<1} $ where $\chi $ denotes the characteristic function. This shows that 
\begin{align*}
\lim_{s\rightarrow + \infty} K(s) =+\infty \;  
\end{align*}
In the same way it is possible to prove that 
\begin{align*}
\lim_{s\rightarrow - \infty} K(s) =+\infty \; , 
\end{align*}
that ensures that $K$ has at least one stationary point. 

As a second step, we find a bound for $s=G_1(\zs)$ which is uniform with respect to $\zs$. Here, we assume that $2n$ (the degree of $p$) is greater than 2 otherwise the result is trivial. Integration by part leads to 
\begin{align*}
p_1(s,\zs) =&
 \frac{\zs^{1/2}}{2\pi^{1/2}}  \int_{-\infty}^\infty \dpt{ p} {x}  e^{-(x-s)^2\zs} \dif x  \;. 
\end{align*}
In the limit of large $s$ we have
\begin{align*}
\dpp{p_1}{\zs} =& -s^{2n-3}  \frac{\dif^{2n} p}{\dif x^{2n}} (0) \frac{ (2n-1)(2n-2)\sqrt{\pi}}{(2n)!\; 2\zs} +o\left(\frac{1}{s}\right) \;, 
\end{align*}
that is obtained by direct computation of the integral. 
Since $\frac{\dif^{2n} p}{\dif x^{2n}} (0)>0$, for $s$ sufficiently large, the derivative of $p_1$ with respect to $\zs$ is negative. Thus, there exists $s_M$ such that  $|s|>s_M$ implies
 \begin{align*}
  p_1(s,\zs) \geq& \lim_{\zs\rightarrow\infty} p_1(s,\zs)
=   \frac{ 1 }{2 }  \dpt{ p}{x}(s) > 0  \;. 
\end{align*}
In conclusion, the solutions $G_1(\zs)$ belong to a bounded set.

Concerning the second constraint, proceeding in the same way, we fix $\overline{s}$ and we prove that there exists $k$ such that the equation $p_2(\overline{s},\zs)=k$ admits a solution $\zs=G_2(\overline{s})$. Integrating by parts, form Eq. \eqref{p2_lemma-proof}, $p_2(\overline{s},\zs)=k$ can be written
\begin{align}
\frac{\sqrt{\zs}}{2}\int^{\infty}_{-\infty} \dst{p}{x} e^{-(x-\overline{s})\zs} \dif x = k\;. \label{comp svil1}
\end{align}
If $p$ is a second order polynomial,  Eq. \eqref{comp svil1} for $k=\frac{\sqrt{\pi}}{2}\dst{p}{x}(0)$ is satisfied for every $\zs >0$.
If the degree of $p$ is  greater than two it is easy to verify that
\begin{align*}
\lim_{\zs\rightarrow \infty}\frac{\sqrt{\zs}}{2}\int^{\infty}_{-\infty} \dst{p}{x} e^{-(x-\overline{s})\zs} \dif x &=\frac{\sqrt{\pi}}{2}\dst{p}{x}(\overline{s}) \\
\lim_{\zs\rightarrow 0}\frac{\sqrt{\zs}}{2}\int^{\infty}_{-\infty} \dst{p}{x} e^{-(x-\overline{s})\zs} \dif x &=+\infty \;.
\end{align*}
If we take $k> \frac{\sqrt{\pi}}{2}\dst{p}{x}(\overline{s}) $, Eq. \eqref{comp svil1} has always a solution. Since we have shown that $\overline{s}$ belongs to a bounded set, there exists $k_M=\sup_{\overline{s}<s_M}k< \infty $. The previous limits show also that we can always find a solution of Eq. \eqref{comp svil1} $\zs\leq \zs_M$ where $\zs_M$ is the solution of Eq. \eqref{comp svil1} for $k=k_M$. The solution $\zs $  belongs to a bounded domain and this concludes the proof of the Lemma.
\hfill $\Box$

\subsection{Macroscopic quantities}\label{sec mean quant}
In order to give the correct physical interpretation to our set of parameters, we investigate the connection of $a_0$, $s$ and $\zs$, with some relevant physical quantities. The conservation of the L$^2$ norm of the wave function $\zy$ provides a simple formula where the coefficient $a_0$ is expressed as a function of $\zs$.
\begin{align}
a_0 =& \pi^{ 1/4} h_0^\zs \int_\mathbb{R}  P (x)  e^{- x^2\zs    }  \dif x=  \pi^{ 1/4} h_0^\zs \int_\mathbb{R} |\zy|^2  \dif x=  \pi^{ 1/4} h_0^\zs \|\zy\|^2_2 = \zs^{1/4} \;.\label{a0}
\end{align}
The mean particle position is given by 
\begin{align*}
\left\langle x \right\rangle  = &  \int_\mathbb{R} x  P (x-s) e^{ -  (x-s)^2 \zs  }\dif x   =  s+ \int_\mathbb{R} x  P (x) e^{ -  x^2 \zs  }\dif x    =  s+ \frac{1}{\pi^{1/4}} \sum_{n=0}^\infty   a_n\int_\mathbb{R} x  h_n^\zs (x)  e^{ -  x^2 \zs  }\dif x  \\
= &     s+ \frac{a_1}{\zs^{ 3/4} \sqrt{2}} \;. 
\end{align*}
Here, angular brackets denote the quantum mechanical expectation value. The particle momentum is given by
\begin{align*}
\left\langle-i  \dpp{}{x} \right\rangle  &= -  i   \int_\mathbb{R} \overline{\zy}(x) \dpp{}{x} \zy (x) \dif x  =  -  i   \int_\mathbb{R} \overline{\zy}(x+s) \dpp{}{x} \zy (x+s) \dif x  \\ 
&=     \int_\mathbb{R}  P \dpp{\chi}{x} e^{ -  x^2 \zs } \dif x=\sum_{n=0}^\infty  \sqrt{2\zs (n+1) } a_{n+1}\chi_n \;.
\end{align*}
where in the last equality we used the expansions of Eqs. \eqref{expans P}-\eqref{expans chi}, the orthogonality property \eqref{ortho h} and Eq. \eqref{der hn-zs}.
A simple way to obtain the evolution of the mean particle position is to apply the Ehrenfest theorem
\begin{align}
\dpt{\left\langle x \right\rangle }{t}   = &  \left\langle- i  \dpp{}{x} \right\rangle \;. \label{HT s_punto}
\end{align}
In our formalism this gives 
\begin{align}
\dot{s}&=  -\frac{1}{\sqrt{2}}  \dpt{}{t} \left( \zs^{-3/4}a_1\right) +\sum_{n=0}^\infty  \sqrt{2\zs (n+1) } a_{n+1}\chi_n \;.  \label{s_punto_meanquant}
\end{align}
This equation agree with Eq. \eqref{dot a1} that will be derived applying the Euler-Lagrange formalism.
We remark that in Eq. \eqref{s_punto_meanquant}, for the sake of generality, we have considered a generic polynomial $P$, for which all the coefficients of the  Hermite expansion may be different from zero. In our case $a_1=0$ by construction and the first term on the right side vanishes. 

\subsection{Quantum Lagrangian}
In this section together with sec. \ref{sec evol eq} we derive the relevant evolution equations for our system. The calculation proceeds straightforwardly. The Euler-Lagrange equations are given by Eq. \eqref{Eu-Lag2}, where the quantum Lagrangian is written in Eq. \eqref{lagr_gen}. We expand the modulus and the phase of $\psi$ on the complete Hermite basis by Eqs. \eqref{expans P}-\eqref{expans chi}. For the sake of generality, in our calculations we assume $a_1\neq 0$ and $a_2\neq 0$. 
We obtain
\begin{align*}
\int_\mathbb{R}   P  \dot{\chi}     e^{- x^2 \zs  }  \dif x =& \sum_{n=0}^\infty a_n \left[\dot{\chi}_n+\frac{\dot{\zs}}{2\zs} \left( \frac{2n+1}{2 } \chi_n +  \sqrt{(n+2)(n+1) } \; \chi_{n+2} \right) \right]\\
\int_\mathbb{R}   P  \dpt{\chi}{x}     e^{- x^2 \zs  }  \dif x =& \sqrt{2\zs} \sum_{n=0}^\infty \sqrt{ (n+1)}  \chi_{n+1}  a_n\\
\frac{\zs^2}{2}  \int_\mathbb{R} x^2P    e^{- x^2 \zs  }  \dif x=& \frac{\zs^{3/4} }{4} \left(\sqrt{2}\; a_0 + a_2\right)\\ 
\int_\mathbb{R}    \frac{x\zs}{2}  \dpt{ P}{x}   e^{- x^2 \zs  }  \dif x
&=\frac{\zs^{3/4}}{\sqrt{2}}\; a_2\;.
\end{align*}
We used the orthogonality condition \eqref{ortho h} and Eq.  \eqref{der zs hn-zs}.
%
Moreover, 
\begin{align}
  \frac{1}{2} \int_\mathbb{R}    P   \left( \dpt{\chi}{x} \right)^2     e^{- x^2 \zs  }  \dif x =&  \pi^{1/4} \zs \sum_{n=0;r,s=1}^\infty\sqrt{ r s} a_n\chi_r \chi_s \int_\mathbb{R}     h_n^\zs  (x) h_{r-1}^\zs   (x)  h_{s-1}^\zs  (x)       e^{- x^2 \zs  }  \dif x \nn \\ 
=&  \pi^{1/4} \zs^{5/4} \sum_{n=0;r,s=1}^\infty\sqrt{ r s} a_n\chi_r \chi_s \int_\mathbb{R}     h_n^1  \left( x  \right)  h_{r-1}^1 \left( x \right)   h_{s-1}^1   \left( x  \right)       e^{- x^2    }  \dif x \;, \label{int p d chi}
\end{align}
where we used Eq. \eqref{scal h zs} and Eq. \eqref{der hn-zs}. We define the  matrix
\begin{align}
A_{r,s,n} =& \pi^{1/4}\int_\mathbb{R}    h_{r}^1  ( x )   h_{s}^1 (x)    h_{n}^1 (x)   e^{   -x^2} \dif x= \sqrt{\frac{2^{-(n+s+r)}}{\pi\, n!s!r!\,} } \int_\mathbb{R}   H_{r}  ( x )   H_{s} (x)    H_{n} (x)   e^{   -x^2} \dif x\;,\label{A def}
\end{align}
where the polynomials $H_n(x)$ are given in Eq. \eqref{H hermite}. Since the matrix $A_{r,s,n}$ is symmetric with respect to the permutation of the indices, it is convenient to assume that the indices $r,s,n,$ are ordered increasingly i. e. $r\leq s\leq n$. The integrals \eqref{A def} can be solved analytically  (see for example \cite{Andrews_book}
\begin{align*}
A_{r,s,n} =& \left\{ 
\begin{array}{ll}
\disp \frac{\sqrt{r!  s! n!  } }{\left(\frac{r+s-n}{2}\right)!\left(\frac{s+n-r}{2}\right)!\left(\frac{n+r-s}{2}\right)!} & \textrm{if }  n-r\leq s\leq n \quad \textrm{and } r+s+n  \textrm{ even} \\[2mm]
0 & \textrm{otherwise }\;. 
\end{array}
\right.
\end{align*}
For later references, we write the first terms 
\begin{align}
A_{0,s,n} =&   \zd_{n,s}\label{A0} \\
A_{1,s,n} =    
&     \sqrt{ s+1}\zd_{n,s+1} +\zd_{n,s-1}\sqrt{s}  \label{A1}  \\
A_{2,s,n} =& \frac{ 1  }{\sqrt{ 2}}  \left[  (\sqrt{ (n+1)(s+1)}+\sqrt{n s}-1) \zd_{n,s} +\sqrt{(s+1)n} \zd_{n-1,s+1}+\sqrt{(n+1)s} \zd_{n+1,s-1} \right]\label{A2}  \;.
\end{align}
At this stage of the calculations, the quantum Lagrangian becomes
\begin{align}
L = &\sum_{n=0}^\infty a_n \left[ \dot{\chi}_n+\frac{\dot{\zs}}{2\zs}   \left( \frac{2n+1}{2 } \chi_n + \sqrt{(n+2)(n+1) } \chi_{n+2} \right)\right] -\dot{s}\; \sqrt{2\zs} \sum_{n=0}^\infty \sqrt{(n+1)}  \chi_{n+1}  a_n \nn\\&+\frac{\zs^{3/4}}{4} \left( \sqrt{2}\; a_2- a_0\right) -  \zs^{5/4} \sum_{n=0;r,s=1}^\infty\sqrt{ r s} a_n\chi_r \chi_s A_{n,r-1,s-1} 
%
 -\sum_{n=0}^\infty   \frac{a_n}{\pi^{1/4} } \left\langle  U_{[s]} , h_n^\zs   \right\rangle_\zs\;+\mathcal{B}, \label{Lagr} 
\end{align}
%
%
where we have defined 
\begin{align}
\mathcal{B}\equiv &    - \frac{1}{8}\int_\mathbb{R}  \frac{1}{P} \left( \dpt{P}{x} \right)^2     e^{- x^2 \zs  }  \dif x\;, \label{bohm_pot_1a} 
\end{align}
and 
\begin{align*}
\left\langle  U_{[s]} , h_n^\zs   \right\rangle_\zs =\int_\mathbb{R} U (s-x) h_n^\zs (x)   e^{- x^2 \zs  }  \dif x \;.
\end{align*}
\section{Bohm potential term}
In this section we analyze into detail some properties of the  term $\mathcal{B}$ that appears in the expression of the quantum Lagrangian. This term  deserves special attention for two reasons. The first reason is related to the computation of $\mathcal{B}$, the presence of $P$ at the denominator may lead to divergences in the computation of the Lagrangian and  numerical instabilities are expected to appear in the numerical treatment of the evolution equations. The second reason is related to the physical interpretation of $\mathcal{B}$. In sec. \ref{sec Bohm pot} we will describe the connection between the term $\mathcal{B}$ and the so called Bohm's potential. 
Bohm observed that the quantum transport is formally equivalent to the evolution of a fluid in which the classical force filed contains an additional term that account for the quantum correction to the motion. Such a force term is expressed by the derivative of the Bohm potential. One of the major limitations of the Bohm representation is related to the highly nonlinear and singular form of the Bohm potential.   
In our formalism, $\mathcal{B}$ coincides with the expectation value of the Bohm potential except for some   smooth contributions. For this reason, the analysis of the behaviors of $\mathcal{B}$ and in particular the study of the divergence of $\mathcal{B}$ in correspondence to some particular values of the parameters $a_n$ and $\zs$, could provide useful informations concerning the role played by the Bohm potential on the quantum motion. Moreover, our method provides a systematic and stable procedure for the numerical approximation of the Bohm potential, and has potentially a large spectrum of applications. We start with the following
\begin{lemma}
	We have
	  \begin{align}
	  \mathcal{B}&   =  - \frac{ \zs^{3/4}}{4}    \sum_{n=1}^\infty  n R_n       a_n   \;, \label{bohm pot}
	  \end{align}
	  where $R_s$ are the expansion coefficients of $\ln (P)$ into the Hermite base
	  . The expansion coefficients of $|\zy|^2$ (Eq. \eqref{exp psiquad}) satisfy the following quasi linear system
	  \begin{align}
	  a_n 
	  =  \sum_{s =1}^\infty B_{n,s}(a)   {R_s}\;, \quad\textrm{with } n\geq1 \;, \label{a_eq_BR}
	  \end{align}
	  %
	  %
	  where the matrix $B$ depends linearly on the coefficients $a_n$ and is given by
	  \begin{align}
	  B_{n,s}(a)=& \sqrt{ \frac{s}{n}   }      \sum_{r =1}^\infty a_r  A_{n-1,r,s-1} \;,  \quad\textrm{with } n\geq1 \;.  \label{B_from_a}
	  \end{align}
\end{lemma}
\emph{Proof.}
We verify the Lemma by direct computation. As a first step, we write $\mathcal{B}$ in the following form
\begin{align}
\mathcal{B}&   =- \frac{1}{8}\int_\mathbb{R} P   \left( \dpt{  R}{x} \right)^2     e^{- x^2 \zs  }  \dif x  \;,\label{bohm_pot_1} 
\end{align}
where we have used that the polynomial $P$ is non negative and we have defined  $
R(x) =  \ln (P(x))$.  Proceeding as in Eq. \eqref{int p d chi}, we obtain
%
\begin{align}
\mathcal{B}&     = -  \frac{ \zs^{3/4}}{4}    \sum_{n=0;r,s=1}^\infty\sqrt{ r s}\;  a_n R_s R_r A_{n,s-1,r-1}\label{B cal0}\;,
\end{align}
where the coefficients $R_s$ are obtained by expanding  $R(x)$ in the Hermite base
$
R(x) = \pi^{1/4} \zs^{-1/4}\sum_n R_n h_n^\zs (x)
$.
 We remember that   $\{a_n\}$ is the set of the expansion  coefficients of $P$ in the Hermite base (see Eq. \eqref{expans P}). Assuming $n\geq1$, we have
\begin{align}
  R_n=&\frac{\zs^{1/4}  }{\pi^{1/4}}   \langle   \ln (P), h_n^\zs\rangle_{\zs} =  \frac{\zs^{1/2}  }{\pi^{1/4}}  \int_\mathbb{R} h_n^1 (x\sqrt{\zs})  \ln \left(\zs^{1/4} \sum_{r=0}^\infty h_r^1 (x\sqrt{\zs}) a_r  \right)  e^{-x^2 \zs} \dif x \nn \\
=&\frac{1  }{\pi^{1/4}} \int_\mathbb{R} h_n^1 (x)  \ln \left( \sum_{r=0}^\infty h_r^1 (x) a_r  \right)  e^{-x^2} \dif x \;.\label{R vs a ditetto}
\end{align}
 By using the elementary properties of the Hermite polynomials, we write the relationship between $\{R_n\}$  and $\{a_n\}$ in  algebraic form. We have
\begin{align*}
a_n =& \pi^{1/4} \int_\mathbb{R}  P (x)  h_n^\zs  e^{- x^2\zs    }  \dif x\nn \\
=&    \pi^{1/4}  \int_\mathbb{R}     e^{- x^2\zs +R(x)   }  h_n^\zs (x)  \dif x= \frac{ \zs^{1/4-\frac{n}{2}}}{\sqrt{2^n n! }} \int_\mathbb{R}     e^{ R (x)   } (-1)^n   \frac{\dif^{(n)}}{\dif x^{n}} e^{-x^2\zs }   \dif x\nn \;,
\end{align*}
where we have used the definition of Hermite polynomials. We proceed by integrating by parts
\begin{align}
a_n =&-     \frac{\zs^{1/4-\frac{n}{2}}}{\sqrt{2^n n! }} \int_\mathbb{R}  \left(\frac{\dif }{\dif x }   e^{    R( x )   } \right) (-1)^n   \frac{\dif^{(n-1)}}{\dif x^{n-1}} e^{-x^2 \zs}  \dif x\nn\\
= & \frac{ \pi^{1/4} \zs^{ -\frac{n}{2}}}{\sqrt{2^n n! }}\sum_{s=1}^\infty  R_s   \int_\mathbb{R}   \frac{\dif  h_s^\zs}{\dif x }        e^{   R ( x )   }  (-1)^{n-1}   \frac{\dif^{(n-1)}}{\dif x^{n-1}} e^{-x^2\zs }  \dif x\nn\\
=&   \frac{\pi^{1/2}  \zs^{-3/4}}{\sqrt{2 n  }} \sum_{s=1}^\infty   R_s\int_\mathbb{R}   \frac{\dif  h_s^\zs}{\dif x }   P(x )    h_{n-1}^\zs (x)   e^{   -x^2\zs } \dif x\nn\\
=&   \frac{ \pi^{1/4} }{\sqrt{  n  }} \sum_{s=1;r=0}^\infty R_s a_r  \sqrt{  s  } \int_\mathbb{R}    h_{s-1}^1  ( x )   h_{r}^1 (x)    h_{n-1}^1 (x)   e^{   -x^2} \dif x= \frac{1}{\sqrt{  n  }} \sum_{s=1;r=0}^\infty R_s a_r  \sqrt{  s  } A_{s-1,r,n-1}\;. \label{an vs Rs} 
\end{align}
In the last line we have used Eq. \eqref{der hn-zs}. Equation \eqref{an vs Rs} coincides with Eqs. \eqref{a_eq_BR}-\eqref{B_from_a}. Finally, inserting Eq. \eqref{an vs Rs} in Eq. \eqref{B cal0}, we obtain Eq. \eqref{bohm pot}.\hfill $\Box$

This Lemma provides an intermediate result that will be used to evaluate the Bohm potential $\mathcal{B} $ in terms of the coefficients $a_n$. Naively, one can calculate the coefficients $R_n$ by solving Eq. \eqref{R vs a ditetto} and  obtain $\mathcal{B}$   by  Eq. \eqref{bohm pot}. However, Eq. \eqref{R vs a ditetto} is not suitable for numerical computation. In fact, the calculation of $R_n$ from Eq. \eqref{R vs a ditetto} requires the integration of highly oscillating Hermite polynomials which becomes a prohibitive task for large values of $n$. Moreover, the numerical calculation of the integral is critical around the points where the polynomial inside the logarithm is small.

The advantage of using the formulation of the problem provided by the Lemma is that whenever the inverse of the matrix $B$ exists, the coefficients $R_s$ are easily obtained by
\begin{align}
R_s =&   \sum_{r =1}^\infty \left[B(a)\right]^{-1}_{s,n}  a_n\;, \quad\textrm{with } s\geq1 \;. \label{R_eq_invBa}
\end{align} 
The proof of the existence of $B^{-1}$ is given in Section \ref{sec inv B}. We write here a useful formula 
 \begin{align}
 - \frac{ \zs^{3/4}}{4}     \dpt{}{a_n}\sum_{n=1}^\infty  n R_n   a_n    =&      2  n      R_n  -   \sum_{r,s\geq 1}^\infty \sqrt{rs} R_sR_r A_{n,r-1,s-1} \label{simp der R_an}\;.
 \end{align}
We give the derivation of this formula in Appendix \ref{sec append Rnan}.

\subsection{Invertibility of the matrix $B(a)$}\label{sec inv B}
We study the invertibility of $B$. This is of primary importance for designing efficient numerical algorithms. 

%
We define the functional spaces that we use to prove the invertibility of the matrix $B$. We define dy $\mathcal{S}^N$ the space of polynomials with degree $N$ and by $\Pi_N$ the projection operator in $\mathcal{S}^N$. We define by $\mathcal{S}_M^N\equiv\left\{P \in \mathcal{S}^N | P=P-\Pi_MP \right\}$ the space of polynomials with degree $N$ that do not contain terms with degree lower or equal to $M$. The space $\mathcal{S}_M^N$ is spanned by the canonical basis set $\left\{x^n \; |\; M \leq n \leq N \right\}$. In particular, for $M=0,1$ the spaces $\mathcal{S}_0^N$ and  $\mathcal{S}_1^N$ are the spans of the following subsets of the Hermite polynomials, respectively $\left\{h_n^\zs \; |\;  n=0,\ldots, N \right\}$ and $\left\{h_n^\zs \; |\;  n=1,\ldots, N \right\}$. 

We denote by $\mathcal{F}_N:\mathbb{R}^{N+1}\rightarrow \mathcal{S}_0^N$ the linear operator that maps the sequence  $\{a_n| n=0,\ldots,N\}$ to the  polynomial $ P (x)= \frac{1}{\pi^{1/4} }\sum_{n=0}^N h_n^\zs (x) a_n$. From the uniqueness of the expansion of a polynomial on the Hermite basis we have that inverse map $\mathcal{F}_N^{-1}$ exists.
We define by $\mathsf{B}(P)\equiv \mathcal{F}_N B \mathcal{F}_N^{-1} :\mathcal{S}_0^N \rightarrow L \left(\mathcal{S}_1^N\right) $  the linear  operator whose matrix representation is given by Eq. \eqref{B_from_a}. With the previous definition, Eq. \eqref{a_eq_BR} is equivalent to 
 \begin{align}
 P(x)= \mathsf{B}(P) R (x)
\end{align}
where $P=\mathcal{F}_N (a_n)$ and $R=\mathcal{F}_N (R_n)$. We prove that $\mathsf{B} (P) $ is invertible by showing that 
\begin{teor}\label{teor invB}
Let $P\in \mathcal{S}_0^N$ be a polynomial such that $P(0)\neq 0$. There exists a surjective operator $\Gamma_{P}\in  L \left(\mathcal{S}_1^N\right)$ such that 
\begin{align}
\mathsf{B} (P) \Gamma_{P} \widetilde{P} =& \widetilde{P}  \quad \forall \;  \widetilde{P} \in \mathcal{S}_1^N\;. \label{teor BGP}
\end{align}	
\end{teor}
\emph{Proof:}  We fix $\widetilde{P}\in\mathcal{S}_1^N$. We prove the existence of the operator $ \Gamma_{P} $  by showing that it is always possible to find a polynomial $R\equiv \Gamma_{P} \widetilde{P} \in\mathcal{S}_1^N $ such that $\mathsf{B} (P) R =  \widetilde{P}$, or equivalently
\begin{align}
\widetilde{a}_n =  \sum_{s =1}^N B_{n,s}(a)   {R_s}\;, \quad\textrm{with } n\geq1 \;,  \label{Beq 2}
\end{align}
where as before $\widetilde{a}_n=\mathcal{F}_N^{-1} (\widetilde{P})$ and $R_n=\mathcal{F}_N^{-1} (R)$. We verify that span$(\Gamma_{P})=\textrm{Dom}(\mathsf{B} (P))=  \mathcal{S}_0^N$. 
Equation \eqref{Beq 2} gives Eq. \eqref{a_eq_BR} for $\widetilde{P} = P $. We  prove the theorem by one intermediate step. We  define a auxiliary polynomial $  G\in \mathcal{S}^{N-1}_0$ and two maps $M_1,M_2$, such that $G=M_1 \widetilde{P}$ and $R=M_2 G$. In this way $\Gamma_{P}=M_2 M_1$ is the composition of $M_1$ and $M_2$. This is illustrated by the following scheme 
\begin{align}
& \overset{ \xrightarrow{\makebox[2cm]{$\Gamma_{P}$}} }{
	 \underset{ \overset{ \in}{\mathcal{S}_1^{N}}  }{\widetilde{P}}  \overset{M_1}{\longrightarrow}   \underset{ \overset{ \in}{\mathcal{S}_0^{N-1}}  }{G}  \overset{M_2}{\longrightarrow}  \underset{ \overset{ \in}{\mathcal{S}_1^{N}}  }{R} 
	} \label{scheme zG}
\end{align}
We define the polynomial $G\equiv M_2^{-1} R \in \mathcal{S}^{N-1}_0$ as follows
\begin{align}
G(x) = &\frac{\pi^{1/4}}{\zs^{1/4}} \sum_{s=1}^{N} \left(\mathcal{F}^{-1}_N R \right)_s  \sqrt{  s  }  h_{s-1}^\zs(x) = \frac{\pi^{1/4}}{\zs^{1/4}} \sum_{s=1}^{N} R_s   \sqrt{  s  }  h_{s-1}^\zs(x) 
\;.\label{G def}
\end{align}
It is immediate to verify that the map $M_2$ is biiective.
By substituting the expression of the matrix $A$ given by Eq. \eqref{A def} in Eq. \eqref{a_eq_BR} we obtain
\begin{align}
\sqrt{  n  }\; \widetilde{a}_n =& \frac{\pi^{1/2}}{\zs^{1/4}}  \sum_{s=1 }^N R_s   \sqrt{  s  }\int_\mathbb{R} P(x)  h_{n-1}^\zs  ( x )   h_{s-1}^\zs (x)      e^{   -x^2 \zs} \dif x \;, \quad\textrm{with } n\geq1 \;. \label{an vs Rn 2}
\end{align}
Equation \eqref{an vs Rn 2} becomes
\begin{align*}
\frac{ \sqrt{  n  }}{\pi^{1/4}}  \; \widetilde{a}_n =&  \left\langle  G(x) P(x)    ,  h_{n-1}^\zs  \right\rangle_\zs\;.
\end{align*}
By using Eq. \eqref{hn-zs3} we write now the left side of the equation as follows 
\begin{align*}
\frac{ \sqrt{  n  }}{\pi^{1/4}}\; \widetilde{a}_n =& \sqrt{  n  } \left\langle   \widetilde{P(x)}, h_n^\zs  \right\rangle_\zs \\
=&  \sqrt{2 \zs }\left\langle   \widetilde{P(x)} x,  h_{n-1}^\zs  \right\rangle_\zs   + \frac{1}{ \sqrt{2 \zs } }\int_\mathbb{R}  h_{n-1}^\zs \left(\dpt{ \widetilde{P(x)}}{x} -2x\zs   \widetilde{P(x)} \right)  e^{   -x^2 \zs} \dif x\\
 =&\frac{1}{ \sqrt{2 \zs } }  \left\langle   \dpt{ \widetilde{P(x)}}{x},  h_{n-1}^\zs  \right\rangle_\zs\;.
\end{align*}
%
We have 
\begin{align}
\left\langle  \left(G  P - \frac{1}{ \sqrt{2 \zs } }   \dpt{  \widetilde{P }}{x} \right), h_{n }^\zs  \right\rangle_\zs =&   0 \quad n=0,\ldots,N-1 \;. \label{GP}
\end{align}
The set of the first $ N$ Hermite polynomials $\{h_n^\zs | n=0,\ldots,N-1  \}$ spans $\mathcal{S}^{N-1}_0$. Equation \eqref{GP} ensures that the product $GP$ is equal to $ \frac{1}{ \sqrt{2 \zs } }   \dpt{ P(x)}{x} $ in the polynomial space  $\mathcal{S}^{N-1}_0$. In order to solve Eq. \eqref{GP} it is convenient to write all the  polynomials  on the canonical basis set $\{x^n | n=0,\ldots,N-1  \}$ of $\mathcal{S}^{N-1}_0$. This is equivalent to expand the polynomials in Taylor series and to compare the coefficients  order by order.
With this, Eq. \eqref{GP} becomes
\begin{align}
 \sum_{k=0}^n  {n \choose k} G^{(k)}(0)  P^{(n-k)}(0)    =  &\frac{ \widetilde{P}^{(n+1)}(0)  }{ \sqrt{2 \zs } }     \quad\textrm{with } n=0,\ldots, N-1\;. \label{gp exp}
\end{align}
For $P (0) \neq 0$ we can solve Eq. \eqref{gp exp} recursively. 
\begin{align*}
G^{(n)} (0) =\left\{
\begin{array}{ll}
    \disp \frac{ \widetilde{P}^{(1)} (0)  }{P (0)  \sqrt{2 \zs } } &\textrm{with } n=0, \\[4mm]
   \disp  \frac{ 1}{ P(0)}  \left[ \frac{ \widetilde{P}^{(n+1)}(0)  }{ \sqrt{2 \zs }   }    -   	\sum_{k=0}^{n-1}  {n \choose k} G^{(k)}(0)  P^{(n-k)}(0)\right]   & \textrm{with } n=1,\ldots, N-1\;. 
\end{array}
\right. 
\end{align*}
In conclusion, we have shown that the map $M_1$ is well defined. 
For each $\widetilde{P}\in \mathcal{S}_1^{N}$, by solving Eq. \eqref{gp exp}, we found a polynomial $G\in \mathcal{S}_0^{N-1}$. The map is one to one since if we fix $G\in \mathcal{S}_0^{N-1}$ we obtain $\widetilde{P}\in \mathcal{S}_1^{N}$ by  Eq. \eqref{gp exp}. 
From the coefficients $G^{(n)} (0)$ we obtain $G$ by the Taylor expansion $G(x)= \sum_{n=0}^{N-1} \frac{ G^{(n)} (0)}{n!} x^n$ and   from Eq. \eqref{G def} we obtain  the coefficients $R_s$ 
	\begin{align*}
%
R_s & = \frac{\zs^{ 1/4}}{\pi^{1/4} \sqrt{  s  }  } 	\left\langle    G  , h_{s }^\zs  \right\rangle_\zs \quad\textrm{with }s=1,\ldots, N\;.  
\end{align*}
According to the scheme \eqref{scheme zG}, we have found a map $\Gamma_P= M_2 M_1$ with span$(\Gamma_P)=\mathcal{S}_1^N$.
 As a final step, we verify that Eq. \eqref{teor BGP} is satisfied.  
We proceed straightforwardly. 
\begin{align*}
\sum_{s =1}^N B_{n,s}(a)   {R_s}= &
  \frac{\zs^{ 1/4}}{\pi^{1/4}  \sqrt{n}  }\sum_{s=1;r =0}^N   a_r  A_{n-1,r,s-1}         \int_\mathbb{R}  G(x)  h_{s-1}^\zs(x) e^{-x^2\zs} \dif x  \;, \quad\textrm{with } n\geq1  \\
 = &  \frac{1 }{   \sqrt{n}  }\sum_{s=1;r =0}^N   a_r   \int_\mathbb{R}     h_{n-1}^\zs  (y) h_{r}^\zs   (y)  h_{s-1}^\zs  (y)         G(x)  h_{s-1}^\zs(x)  e^{- (y^2 +x^2)\zs} \dif x  \dif y   \\
 =  &   \frac{\pi^{1/4} }{   \sqrt{n}  }\sum_{s  =1}^N  \int_\mathbb{R}    P(y)     h_{n-1}^\zs  (y)  h_{s-1}^\zs  (y)         G(x)  h_{s-1}^\zs(x)  e^{- (y^2 +x^2) \zs} \dif x  \dif y  \;. 
\end{align*}
%
We use that for each $G\in \mathcal{S}_0^{N-1}$
\begin{align*}
   &    \sum_{s =1}^N          h_{s-1}^\zs  (y)       \int    G(x)  h_{s-1}^\zs(x)  e^{  -x^2\zs} \dif x  =  G(y)\;.
\end{align*}
By Eq. \eqref{GP} 
\begin{align*}
\sum_{s =1}^N B_{n,s}(a)   {R_s}= &   \frac{\pi^{1/4}  }{   \sqrt{n}  }   \int_\mathbb{R}    P(y)     G(y)    h_{n-1}^\zs  (y) e^{- y^2 \zs  }   \dif y   =  \frac{\pi^{1/4} }{   \sqrt{n}  } \left\langle  P      G , h_{n-1}^\zs  \right\rangle_\zs\\
=&\frac{\pi^{1/4} }{   \sqrt{n}  \sqrt{2 \zs }}   \left\langle    \dpt{  \widetilde{P(x)}}{x} , h_{n-1}^\zs  \right\rangle_\zs =\widetilde{a_n}\;.
\end{align*}
This concludes the proof of Theorem \ref{teor invB}.\hfill $\Box$\\
As a final remark, we note that we apply this theorem to a polynomial of the form $P(x-s)$ where the spatial variable is shifted by $s$. Due to the arbitrarily of $s$, the condition $P(0)\neq 0$ implies $P\neq 0$ always. 

\subsection{Calculation of the coefficients $R_n$: analytic approach}

The main source of numerical instabilities of our model of quantum transport arises from the presence of the Bohm term in the Lagrangian. In particular, it is challenging to express the  coefficients $R_i$ that appear in Eq. \eqref{bohm pot} in terms of the dynamical parameters $a_i$.
When the number of parameters is small, it is possible to compute directly $R_i$ as a function of $a_i$ by a simple analytical formula.
%
%
At first, we consider a simple test case whose analysis leads to interesting considerations concerning the singularities showed by the evolution equations of the parameters. We consider a solution containing only two parameters, namely $a_0$ and $a_3$. 
\begin{align*}
a=& (a_0,0,0,a_3)\;. 
\end{align*}
The cutoff index in the expansion on Hermite polynomials is $N=3$. We solve by hand the system
\begin{align*}
R_s =&   \sum_{r =1}^N \left[B(a)\right]^{-1}_{s,n}  a_n\;. \quad\textrm{with } s\geq1 \\
B_{n,s}(a)=& \sqrt{ \frac{s}{n}   }      \sum_{r =1}^N a_r  A_{n-1,r,s-1} \;,  \quad\textrm{with } n\geq1   
\end{align*}
It is convenient to define the matrix $\widetilde A^r_{n,s} \equiv A_{n-1,r,s-1} $. Simple computations lead to (see Eq. \eqref{A def})
\begin{align*}
\widetilde A^0=& \left(\begin{array}{ccc}
1 & 0 &0 \\
0 & 1 &0 \\
0 & 0 &1 \\
\end{array}\right) ; \quad \widetilde A^3= \left(\begin{array}{ccc}
0 & 0 & \sqrt{3} \\
0 &0 &0 \\
\sqrt{3} & 0 &0 \\
\end{array}\right)
\end{align*}
and 
\begin{align*}
%
B (a)=&    \left(\begin{array}{ccc}
a_0 & 0 & 3 a_3 \\
0 &a_0 &0 \\
a_3 & 0 &a_0 \\
\end{array}\right)\;.
\end{align*}
We are interested only on the coefficient $R_3$. Cramer's rule provides 
\begin{align}
R_3 =&  \frac{a_0 a_3}{a_0^2 - 3 a_3^2}\;.
\label{R anal1}  
\end{align}
In order to show how of the complexity of the method increases when other coefficients beyond $a_0$ and $a_3$ are considered, we give the results of the computation for the case $N=4$. 
We proceed in the same way as before and we obtain  the coefficients $R_i$ with $i=1,\ldots,4$  
\begin{align}
R_1&=a_0( 3 \sqrt{6}a_3^3 -2a_0 a_3a_4  -2\sqrt{6} \;  a_3a_4^2  )/|B|\label{R anal2a}\\
R_2&=(3\sqrt{2}  a_3^4 -3 \sqrt{2}  a_0^2a_3^2 -4 \sqrt{2}a_0^2a_4^2 + 6 \sqrt{3} a_0 a_3^2a_4 - 8 \sqrt{3} a_0 a_4^3)/(2|B|)\\
R_3&=a_0( a_3 a_0^2 +\sqrt{6} a_3 a_0 a_4 - a_3^3)/|B|\\
R_4&= a_0^2 (2a_0a_4 +2\sqrt{6}a_4^2
-  3\sqrt{6}a_3^2 )/
(2|B|) \label{R anal2b} \\
|B|&=a_0^4 + 3a_3^4 - 22a_0^2a_3^2 + 14a_0^2a_4^2 - 4\sqrt{6}a_0a_4^3 + 4\sqrt{6}a_0^3a_4  + 2\sqrt{6}a_0a_3^2a_4\;.
\end{align}
The formulas that express the coefficients $R_i$ from the parameters $a_i$ become cumbersome when $N$ increases. For this reason, in order to improve the precision of the calculations it is necessary to resort to numerical methods. The most convenient way, is to calculate the matrix $B$ from Eq. \eqref{B_from_a} and to obtain $B^{-1}$ numerically. Theorem \ref{teor invB} ensures that this is always possible.

For the sake of clarity, we summarize the equations and the link among the variables in the following scheme
\begin{align*}
\begin{array}{lllll}
I:&  a_n&  \xrightarrow{\makebox[1cm]{\scriptsize Eq. \eqref{expans P}}}  &  P = \pi^{-1/4} \sum_{n=0}^\infty h_n^\zs a_n  \xrightarrow{\makebox[1cm]{\scriptsize Eq. \eqref{R vs a ditetto} }} R_n =   \zs^{1/4}\pi^{-1/4}    \langle   \ln (P), h_n^\zs\rangle_{\zs} \\[2mm]
II:  & a_n & \xrightarrow{\makebox[1cm]{\scriptsize Eq. \eqref{R anal1} }}   &    R_3 =   \frac{a_0 a_3}{a_0^2 - 3 a_3^2}  \textrm{ ; or Eqs. \eqref{R anal2a}-\eqref{R anal2b}}   \\[2mm] 
III:& a_n & \xrightarrow{\makebox[1cm]{\scriptsize Eq. \eqref{R_eq_invBa} }}   &  R_s=     \sum_{ r} \left[B(a)\right]^{-1}_{s,r}  a_r \;.
\end{array}
\end{align*}
We indicate three possible schemes, in the first case (denoted by $I$) we start by the coefficients $a_n$ we obtain the polynomial $P(x)$, and we obtain the coefficients $R_i$ by projecting the function $\ln P(x)$ on the Hermite polynomials. In principle, this procedure provides the coefficients $R_i$ with greater accuracy. However, the scheme $I$ has little practical use since the projection procedure at the end is very expensive from a computational point of view. The other two cases have been already discussed into details, in the case $II$ we make use of the analytical approximations and  the case $III$ requires basically only the inversion of the matrix $B$.     

\begin{figure}[!h]
	\begin{center}
		\includegraphics[width=0.49\columnwidth]{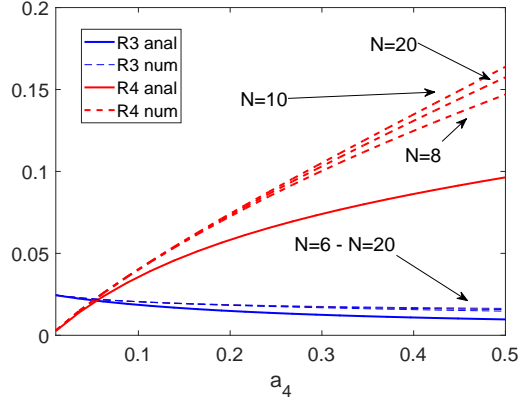}
		\caption{Coefficients $R_3$ (blue line)  and $R_4$ (red line) as a function of $a_4$. Solid lines refer to the Eqs. \eqref{R anal2a}-\eqref{R anal2b} and dashed lines refer to the numerical results for different values of $N$. \label{fig_R} }
	\end{center}
\end{figure}

\begin{figure}[!h]
	\begin{center}
		\includegraphics[width=\columnwidth]{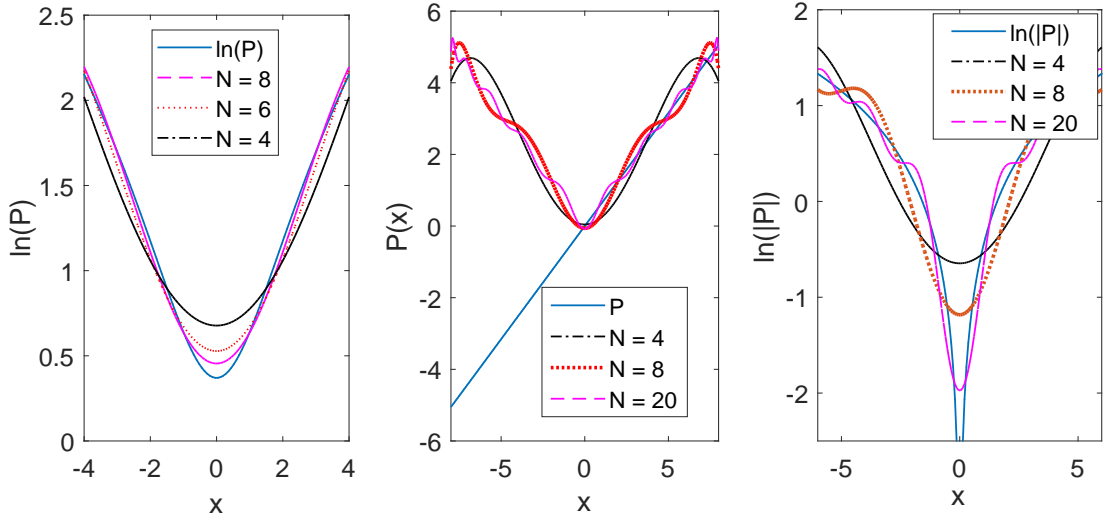}
		\caption{First panel:  $R(x)$ calculated up to the order $N$, the solid line represents $\ln(P)$. Second panel: Plot of $P(x)$ (solid line) used for the calculation of $R(x)$ depicted in the second panel. The dashed lines represent $P(x)$ reconstructed from $R_i$. Third panel: $R(x)$ calculated up to the order $N$, the solid line represents $\ln(|P|)$.  \label{fig_exp_R} }
	\end{center}
\end{figure}

We investigate the precision of different schemes with two examples. 
As a first case, we consider the following polynomial 
\begin{align*}
P(x)=\frac{1}{\pi^{1/4}} \left(2\;h_0^\zs(x)  
+ 0.05\; h_3^\zs(x) +a_4\; h_4^\zs(x) \right) \;.
\end{align*}
We vary the parameter $a_4$ in the interval $[0,0.5]$ and we evaluate the coefficients $R_3$ and $R_4$ by the schemes $II$ and $III$. We use $\zs=1/2$, however the results are independent from the value of $\zs$. The results are depicted in Fig. \ref{fig_R}. In the blue (red) line we depict the value of $R_3$ ($R_4$). We compare the results obtained by using the analytical formulas \eqref{R anal2a}-\eqref{R anal2b} (solid lines) with the values obtained by numerical solution (dashed lines) for different values of the  expansion cutoff $N$ which indicates the precision of the calculation. We find good agreement between the numerical results and the analytical formulas. 
As a second example, we consider the following positive polynomial 
\begin{align*}
P(x)=\frac{1}{\pi^{1/4}} \left(3\;h_0^{1/2}(x)  
+ h_2^{1/2}(x) \right) \;.
\end{align*}
which corresponds to the vector $a=(3,0,1)$ with $N=3$ and $\zs=0.5$. In  Fig. \ref{fig_exp_R} (first panel from the left) we depict the projection of $R(x)=\log (P)$ on the Hermite polynomials for different values of the cutoff $N=4,6,8$. 
\begin{align*}
R(x) =  \sum_{n=1}^N \langle   \ln (P), h_n^\zs\rangle_{\zs} h_n^\zs (x)\;.
\end{align*}
We see that we have a good convergence to the exact result already for $N=8$. In the second and third panels we show the results of the same procedure for a non-positive polynomial ($a_1=1$)
\begin{align*}
P(x)=\frac{ h_1^{1/2}(x)}{\pi^{1/4}} =\frac{1}{2^{1/4}\pi^{1/2}}\;x\;.
\end{align*}
We remark that this is an artificial example, since in our model $P(x)$ represents the squared of the particle density, and thus is necessarily positive. In the second panel of Fig. \ref{fig_exp_R} we plot the polynomial $P(x)$ obtained from  $R_i$. Interestingly, we note that our algorithm converges toward the  absolute value of $P(x)$. This is a very useful property of our method. In fact the cutoff in the Hermite expansion of $P(x)$ necessarily lead to approximations for which the positivity of the particle density is not longer guarantee. The convergence of the algorithm to the absolute value of $P(x)$ is a good property that  prevents that such errors may have catastrophic consequences on the dynamics of the quantum particle.  

\section{Evolution equations}\label{sec evol eq}
We complete the expression of the quantum Lagrangian by using  Eq. \eqref{bohm pot} in Eq. \eqref{Lagr}
\begin{align*}
L = &\sum_{n=0}^\infty a_n \left[ \dot{\chi}_n+\frac{\dot{\zs}}{2\zs}   \left( \frac{2n+1}{2 } \chi_n + \sqrt{(n+2)(n+1) } \chi_{n+2} \right)\right] -\dot{s}\; \sqrt{2\zs} \sum_{n=0}^\infty \sqrt{(n+1)}  \chi_{n+1}  a_n \nn\\&+\frac{\zs^{3/4}}{4} \left( \sqrt{2}\; a_2- a_0\right) -  \zs^{5/4} \sum_{n=0;r,s=1}^\infty\sqrt{ r s}\; a_n\chi_r \chi_s A_{n,r-1,s-1}  \nn\\
&   - \frac{ \zs^{3/4}}{4}    \sum_{n=1}^\infty  n R_n       a_n    -\sum_{n=1}^\infty   \frac{a_n}{\pi^{1/4} }\left\langle  U_{[s]} , h_n^\zs   \right\rangle_\zs \; . 
\end{align*}
The Euler-Lagrange equations $\dpt{}{t} \frac{\partial L}{\partial \dot{a_n} } =  \frac{\partial L}{\partial a_n } $  and $\dpt{}{t} \frac{\partial L}{\partial \dot{\chi_n} } = \frac{\partial L}{\partial \chi_n } $, lead to respectively
\begin{align}
\dot{\chi}_n = & -\frac{\dot{\zs}}{2\zs} \left( \frac{2n+1}{2 } \chi_n + \sqrt{(n+2)(n+1) } \chi_{n+2} \right) +  \dot{s}    \sqrt{2\zs(n+1)}  \chi_{n+1} + \zs^{3/4}\left(\frac{\zd_{n,2}}{2\sqrt{2}} -\frac{\zd_{n,0}}{4}\right)  \nn \\
& - \sum_{r,s\geq 1} \left(\zs^{5/4}  \chi_r \chi_s   - \frac{\zs^{3/4}}{4}R_rR_s \right)\sqrt{rs} A_{n,r-1,s-1}-     \frac{ \zs^{3/4} }{2} nR_n + \frac{1}{\pi^{1/4}} \left\langle  U_{[s]} , h_n^\zs   \right\rangle_\zs   \;,   \label{chi eq0}
\end{align}
and
\begin{align}
\dot{a}_n  = & \frac{\dot{\zs}}{2\zs}  \left(  a_n\frac{2n+1}{2 }      +    a_{n-2}    \sqrt{n(n-1) }   \right) -   \dot{s}  \sqrt{2 n \zs}   a_{n-1} + 2 \zs^{5/4}\sqrt{ n  } \sum_{m, s\geq1}\sqrt{  s}\;  a_m  \chi_s A_{m,n-1,s-1} \label{a eq0}\;.
\end{align}
In order to derive Eq. \eqref{chi eq0} we have used  Eq. \eqref{simp der R_an}. The evolution equation for the spatial coordinate $s(t)$ and for the inverse of the variance $\zs(t)$ are obtained from the evolution equation \eqref{a eq0}, respectively, for $n=1$ and $n=2$. By using Eq. \eqref{A0} we have
\begin{align*}
%
\dot{a}_2 =&  -2\dot{s} \sqrt{\zs} a_1  +  \frac{5}{4 } \frac{\dot{\zs}}{\zs} a_2 + \frac{\sqrt{2 }}{2}\frac{\dot{\zs}}{\zs}   a_0      +2\sqrt{ 2 }  \zs^{5/4} \sum_{n=0}^\infty \left(n a_n \chi_n + \sqrt{n+1}\sqrt{ n+2} a_n \chi_{n+2}\right)  \;. 
\end{align*}
According to the discussion of Sec. \ref{sec comp exp} we set $ a_1 =0$ and $a_2$ constant given by the initial condition $a_2(t_0)$. We obtain the equation for the variable $\zs$
\begin{align*}
 \dot{\zs}        =&   - \frac{16  \zs^{2}}{4+5\sqrt{2} \zs^{-1/4}a_2(t_0)} \sum_{n=0}^\infty a_n \left(n  \chi_n + \sqrt{n+1}\sqrt{ n+2}   \chi_{n+2}\right)  \;,
\end{align*}
where we used $a_0= \zs^{1/4}$. Concerning the variable $a_1$, the evolution equation is
\begin{align}
\dot{a}_1   = &  -\dot{s} \sqrt{2\zs}   a_0  + \frac{3}{4 }   \frac{\dot{\zs}}{\zs}   a_1   +2 \zs^{5/4} \sum_{n=0}^\infty \sqrt{   n+1} a_{n} \chi_{n+1}   \;. \label{dot a1} 
\end{align}
Imposing $a_1=0$
\begin{align*}
\dot{s}  = &    \sqrt{2 }  \zs^{1/2} \sum_{n=0}^\infty \sqrt{   n+1} a_{n} \chi_{n+1}      \;.
\end{align*}
Finally, the evolution equations are
\begin{align}
\dot{\chi}_n = &M\left( \frac{2n+1}{2 } \chi_n + \sqrt{(n+2)(n+1) } \chi_{n+2} \right)   +  2 \zs S\sqrt{n+1}    \;  \chi_{n+1}\nn  +\frac{\zs^{3/4}}{2\sqrt{2}}\left( \zd_{n,2}  -\frac{\zd_{n,0}}{\sqrt{2}}\right)  \nn \\
&-\frac{ \zs^{3/4} }{2} n\; R_n - \zs^{3/4} \sum_{r,s\geq 1}\sqrt{rs} A_{n,r-1,s-1} \left(\zs^{1/2}  \chi_r \chi_s   - \frac{R_rR_s}{4} \right) 
+  \pi^{-1/4}\left\langle  U_{[s]} , h_n^\zs   \right\rangle_\zs \label{chi eq}\\
\dot{a}_n  = & -M \left(  a_n\frac{2n+1}{2 }   +    a_{n-2}    \sqrt{n(n-1) }   \right)- 2 \zs S \sqrt{ n }\;    a_{n-1}  \nn \\
& +2 \zs^{3/2}n\;  \chi_n  + 2 \zs^{5/4}\sqrt{ n  } \sum_{m, s\geq1}\sqrt{  s}\;  a_m  \chi_s A_{m,n-1,s-1}\label{an eq}\\
\dot{\zs}        =&  -  2 \zs M \label{zs eq} \\
\dot{s}  = &    \sqrt{2 }  \zs^{1/2} S \;.\label{s eq}
\end{align}
We have defined
\begin{align}
S  = &   \zs^{1/4} \chi_{ 1}+ \sum_{r=1}^\infty \sqrt{   r+1}\;  a_{r} \chi_{r+1}    \\
M  = &    \frac{8 \zs}{4+5\sqrt{2} \zs^{-1/4}a_2(t_0)} \left[  \sqrt{2} \zs^{1/4}   \chi_{2}  +   \sum_{r=1}^\infty a_r \left(r  \chi_r + \sqrt{(r+1)(r+2)}   \chi_{r+2}\right)\right] \;.
\end{align}
Concerning the coefficient $a_0$, Eq.  \eqref{an eq} with $n=0$ combined with Eq. \eqref{zs eq} gives
\begin{align*}
\frac{\dot{a}_0 }{a_0}  = & \frac{1}{4}  \frac{\dot{\zs}}{\zs} 
\end{align*}
whose solution is $a_0(t)=\zs^{1/4}(t)$, which, as already seen in Sec. \ref{sec mean quant}, ensures that the norm of the wave function is conserved. 

At first, we derive the evolution equation the Gaussian ansatz given in Eq. \eqref{expans chi} as a particular case of the evolution equations for the complete set of parameters. We express the first three coefficients $\chi_n$ by the variables $\phi_0$, $p$ and $\zs_i$. This can be done by considering  the second order polynomials 
\begin{align*}
\pi^{1/4}\sum_{n=0}^2 \chi_n h_n^\zs (x)= & \phi_0 -px + \zs_i\frac{x^2}{2}\;.
\end{align*}
We obtain $
\chi_2 =   \frac{1}{2\sqrt{2}}\zs^{-5/4}\zs_i$, $
\chi_1=-	 \frac{1}{\sqrt{2}}p \zs^{-3/4}$, 
$\chi_0 =   \zs^{-1/4}\phi_0+   \frac{1}{4}\zs^{-5/4}\zs_i$. 
Direct computation shows  that Eq. \eqref{zs eq} reduces to Eq. \eqref{ex zs dot} and Eq. \eqref{chi eq} for $n=0$ and $n=2$ gives, respectively, Eq. \eqref{ex p dot} and Eq. \eqref{ex zsi dot}.

The harmonic oscillator $U (x) = \frac{\zo^2 x^2 }{2} $ is a standard example with several applications. In our model the external potential enters only in the equations for the variables $\chi_n$ via the scalar product with the Hermite polynomials. For the harmonic potential this is easily calculated. We have
\begin{align}
   \frac{1}{\pi^{1/4}}\left\langle  U_{[s]} , h_2^\zs   \right\rangle_\zs=\frac{1}{\pi^{1/4}}\int_\mathbb{R} U (s-x) h_2^\zs (x)   e^{- x^2 \zs  }  \dif x=&
      \frac{ \zo^2\zs^{1/4}}{2\sqrt{ 2} \; \pi^{1/4}}   \int_\mathbb{R}   (s-x)^2   \left( 2\zs x^2-1\right)   e^{- x^2 \zs  }  \dif x \nn\\
      =&  \frac{ \zs^{-5/4}}{2\sqrt{ 2}} \zo^2 \;. \label{har pot h2}
\end{align}
\section{Examples}
We illustrate our method by considering a simple case. In the expansion of the  wave function, we keep one coefficient ($\chi_2$) for the phase and two coefficients  ($a_0, a_3$) for the modulus. 
It is more convenient to use the parameter $\zs_i$ that appears in the Gaussian ansatz instead of $\chi_2$. The change of variable  is given by the  equation  $\chi_2 =   \frac{\zs^{-5/4}}{2\sqrt{2}}\zs_i$. Concerning the potential, we consider the harmonic trap  $U (x) = \frac{\zo^2 x^2 }{2} $. 
We obtain the following reduced system of equations 
\begin{align}
\dot{a}_3  = &-\frac{7}{2 }  \zs_i  a_3 \label{a3dot}\\
\dot{\zs}_i= &    \zs^2 -\zs_i^2 -\zo^2   -   \frac{3}{4}  \frac{ \zs^{2} }{\left(1-\frac{\zs^{1/2}}{3a_3^2} \right)}  \label{sidot}\\
\dot{\zs} = &  -  2  \zs\zs_i  \;, \label{sdot} 
\end{align}
where we have used Eq. \eqref{R anal1} with $a_0=\zs^{1/4}$ and   Eq. \eqref{har pot h2}.
Comparing with Eqs. \eqref{ex zs dot}-\eqref{ex zsi dot}  we see that the only difference is given by the last term of Eq. \eqref{sidot}. In the limit $a_3(t_0) \rightarrow 0$ this term goes to zero and the system reduces to the evolution of a simple Gaussian beam.
In this case $(a_3=0)$ it is convenient to represent the solution on the $\zs_i - \zs$ coordinate plane.  The trajectories followed by the parameters $\zs_i(t)$ and $ \zs(t)$ are circles with centre $(0,\frac{\xi}{2})$ and radius $\sqrt{\frac{\xi^2}{4}-\zo^2}$ where $\xi =\frac{\zs_i^2 (t_0)+\zs^2 (t_0)+\zo^2}{\zs (t_0)}$ and $t_0$ is the initial time. 
From Eq. \eqref{a3dot} and \eqref{sdot} we obtain 
\begin{align*}
\dpt{\zs}{a_3 } = &\frac{4}{7 } \frac{\zs}{a_3}   %
\end{align*}
and 
\begin{align}
a_3 (t)= \frac{a_3(t_0)}{\zs^{\frac{7}{4}}(t_0)}  \zs^{\frac{7}{4}}(t)\label{sol_anal_a3}\;.
\end{align}
\begin{figure}[!t]
	\begin{center}
		a)\includegraphics[width=0.48\columnwidth,height=0.4\textwidth]{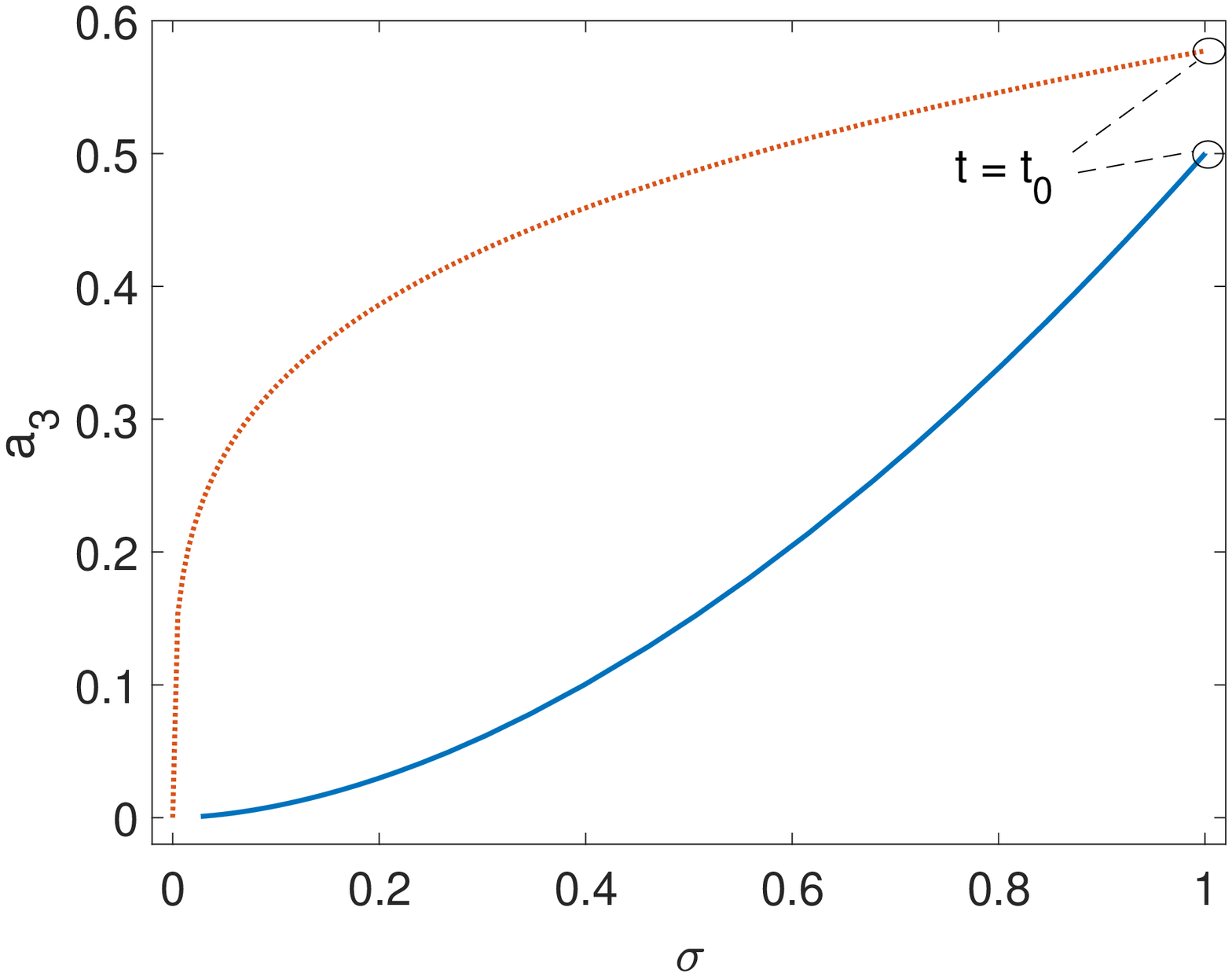}
		\includegraphics[width=0.48\columnwidth,height=0.4\textwidth]{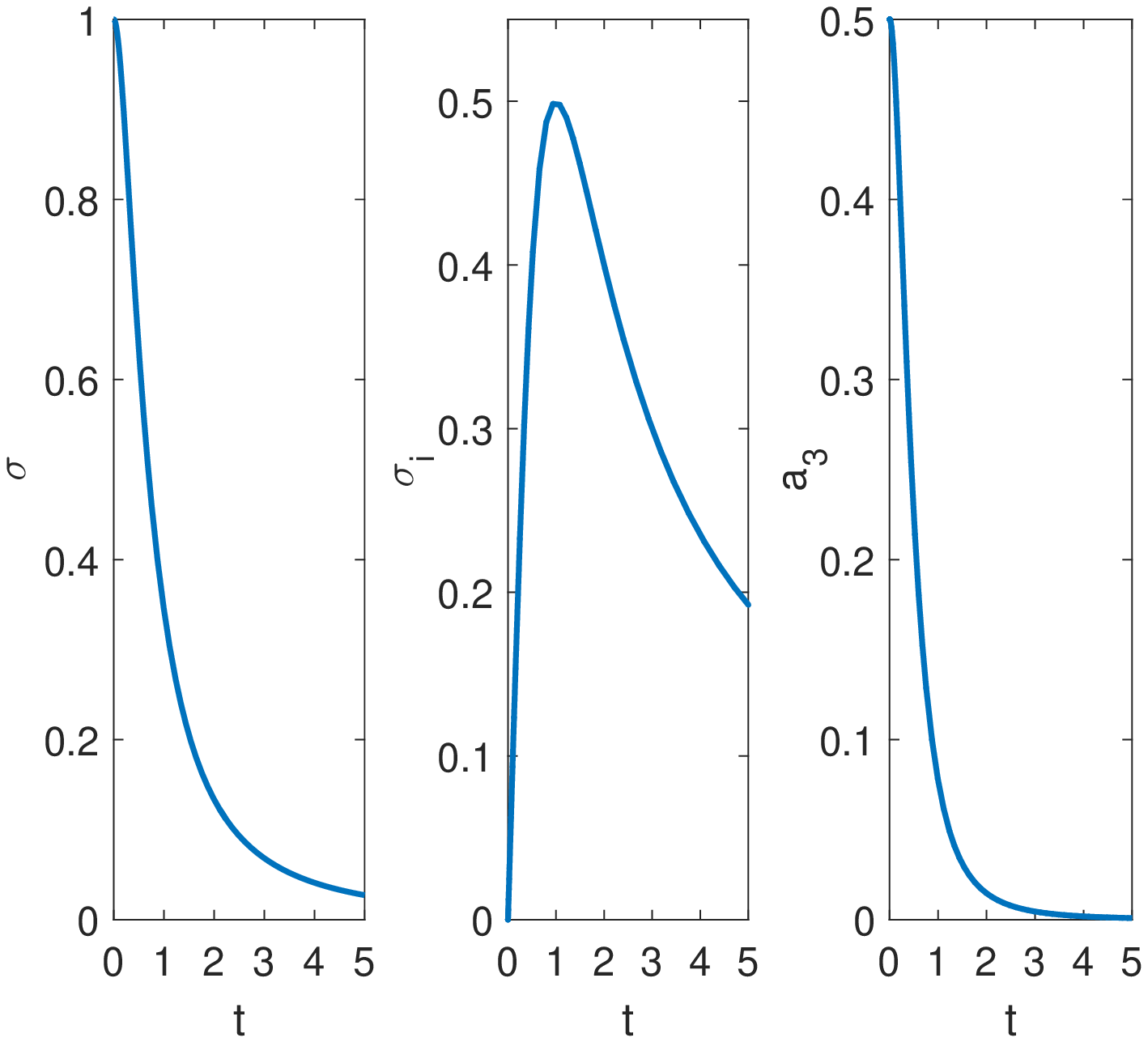}
		b)\caption{Left panel: Trajectory of the solution in the plane $(a_3,\zs)$ for $\zo=0$. The red curve depicts the pole $\zs^{1/2}-3a_3^2 =0 $. Right panel: time evolution of  the variables. \label{fig_a3_vs_sr_1} }
	\end{center}
\end{figure}
We interpret the  last term in the right of Eq. \eqref{sidot}  as the first correction to the Gaussian motion induced by the Bohm potential term. As is often the case in Bohm dynamics, the quantum potential shows singular points. It is important to verify that Eq. \eqref{sidot} is always well defined ($\zs^{1/2}-3a_3^2 \neq 0 $). This is illustrated in Fig.  \ref{fig_a3_vs_sr_1}. We proof that this is always the case. 
Inserting Eq. \eqref{sol_anal_a3} in Eqs. \eqref{sidot}-\eqref{sdot} we obtain
\begin{align}
\dot{\zs}_i= & - \zs_i^2 -\zo^2     +\zs^2 \left(1 -   \frac{3}{4}  \frac{1 }{\left(1-\frac{1}{3\zs^{3} \zh_0^2} \right)} \right) \label{zsidot ex fin} \\
\dot{\zs} = &   - 2  \zs\zs_i \label{zsdot ex fin}  
\end{align}
where we have defined $\zh_0 = \frac{a_3(t_0)}{\zs^{\frac{7}{4}}(t_0)}   $.

\begin{figure}[!h]
	\begin{center}
		\includegraphics[width=0.32\columnwidth]{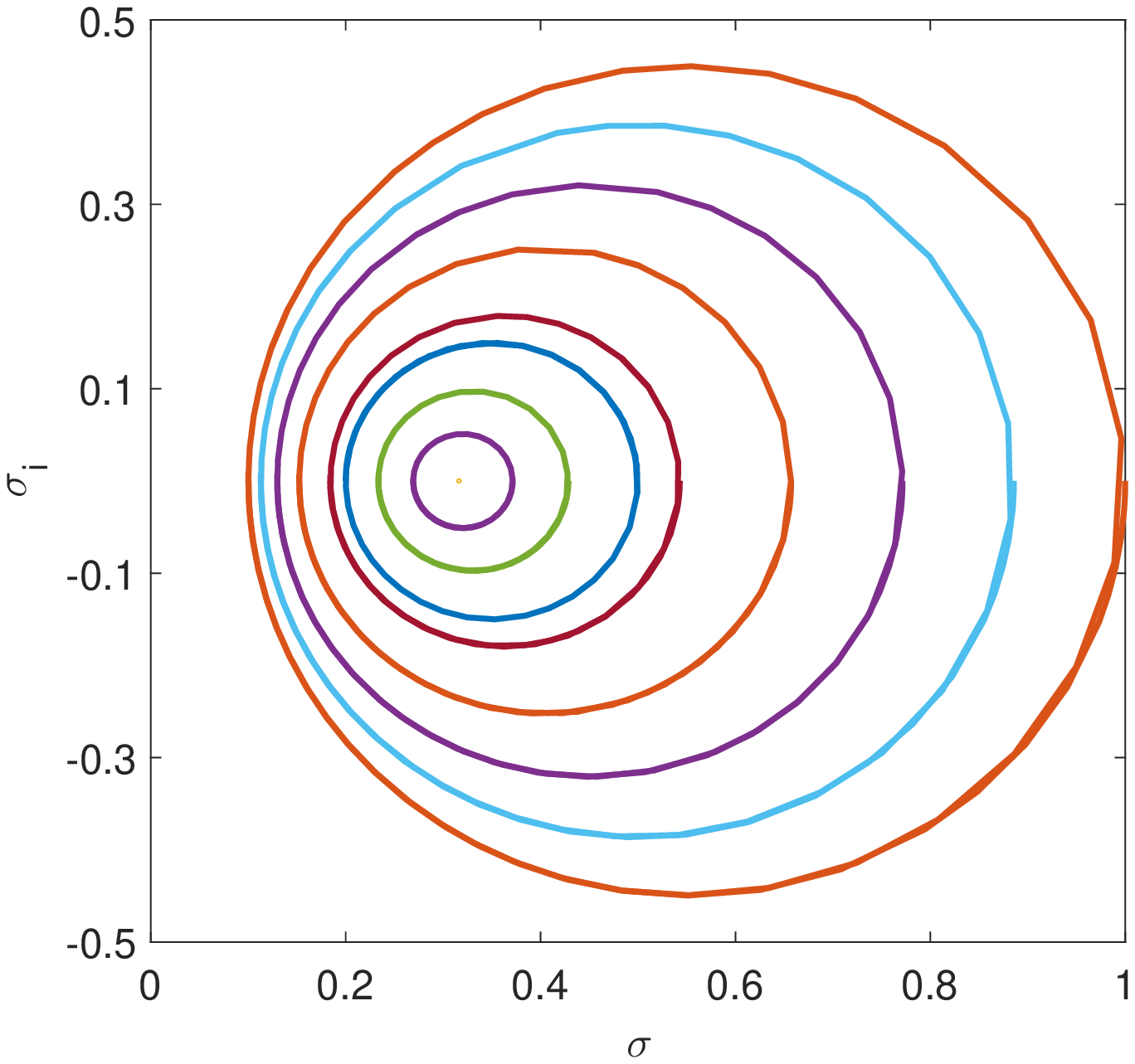}
		\includegraphics[width=0.32\columnwidth]{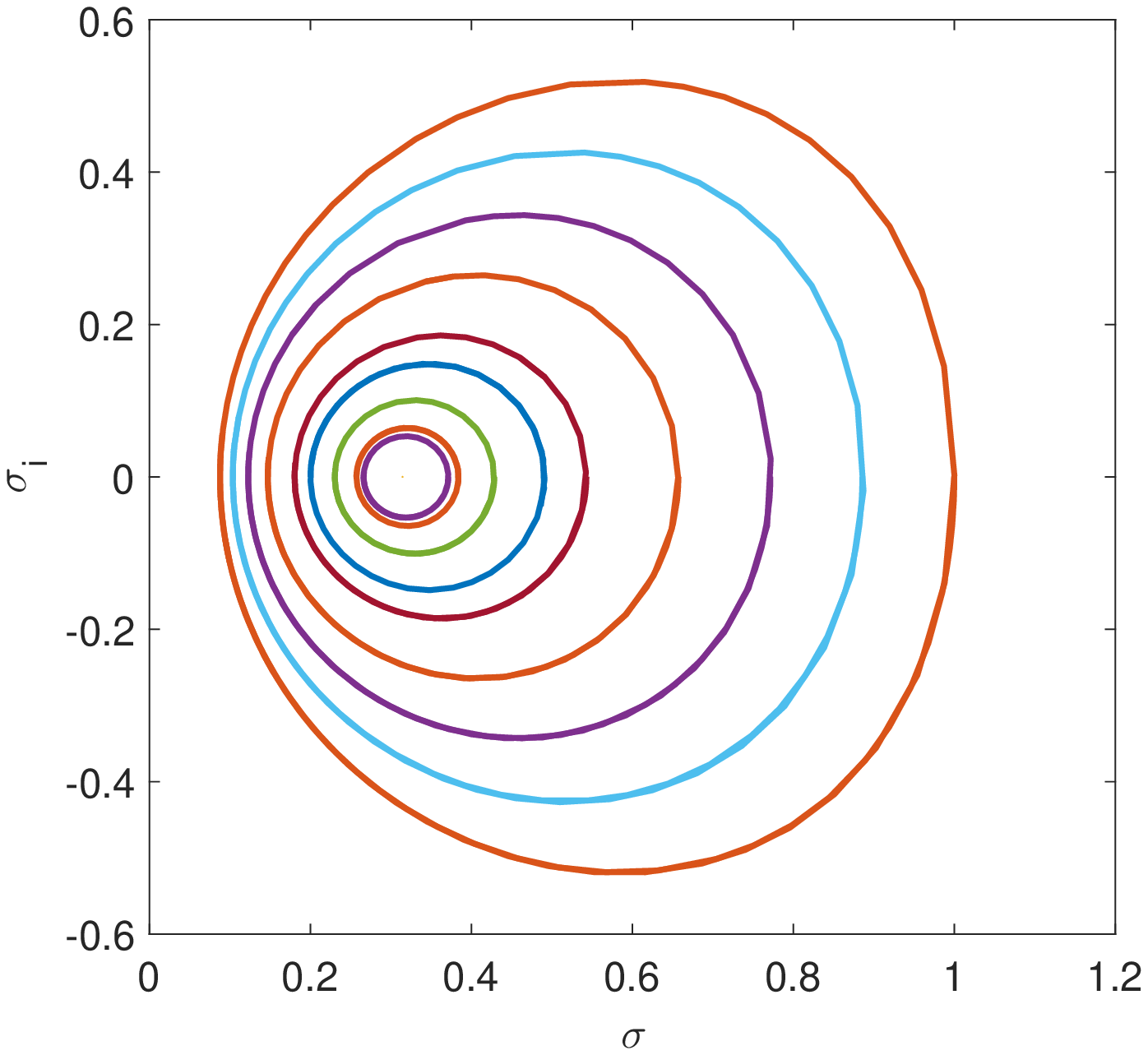}
		\includegraphics[width=0.32\columnwidth]{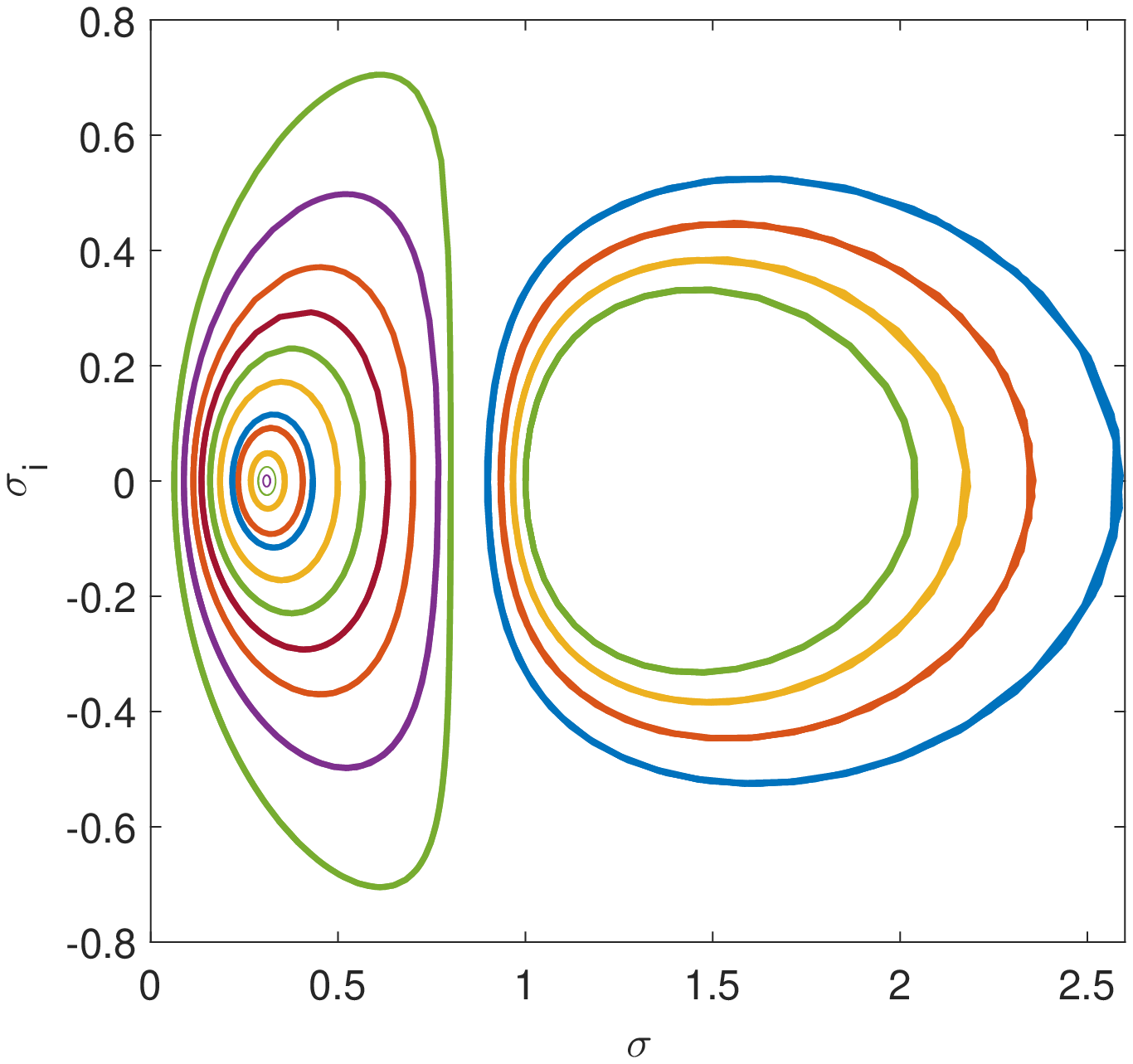}
		\caption{Trajectory of the solution of Eqs. \eqref{zsidot ex fin}-\eqref{zsdot ex fin} in the plane $(\zs,\zs_i)$ for different initial conditions. From the left, the panels depict the solution for increasing values of the parameter $\zh_0$. We used $\zh_0=0,0.4,0.8$ and $\zo=0.1$. \label{fig_zs_vs_sr_1} }
	\end{center}
\end{figure}
In oder to illustrate the behaviors of the solution of Eqs. \eqref{zsidot ex fin}-\eqref{zsdot ex fin}, in Fig. \ref{fig_zs_vs_sr_1} we plot the trajectories of the solution on the $\zs_i$-$\zs$-coordinate plane for different initial conditions. We focus on the deformation of the trajectories caused by the Bohm potential term. For $\zh_0=0$ we obtain the simple Gaussian case and the trajectories are circles (Fig. \ref{fig_zs_vs_sr_1} panel on the left). For $\zh_0\neq 0$ the vertical line $ \zs_0 = \frac{\zh_0^{-2/3}}{ \sqrt[3]3 }$ separate the trajectories into two groups which remain in the left or in the right side of the semi-plane. In particular, our results show that as a function of the initial conditions, the solution may oscillate with two different frequencies.

It is interesting to analyze more into details the behavoiur of the solution close to the  singularity. In particular, we show that the trajectories do not cross the line $ 1-3\zs^{3} \zh_0^2=0$. It is convenient to introduce the variables $y=-\zs_i$ and $z= (1-3\zs^{3} \zh_0^2)^{1/3}$.
In order to study the behaviour of the system around the singularity $z=0$ we introduce a small parameter $\ze$ and we choose as initial condition a point that approaches the axis $z=0$ when $\ze$ goes to zero. In the new variables, the system of Eq. \eqref{zsidot ex fin}-\eqref{zsdot ex fin} becomes
\begin{align}
\left\{
\begin{array}{l}
\disp  \dot{y} =   y^2 +\zo^2     -  \left(\frac{z^3-1}{3 \zh_0^2}\right)^{2/3}  \frac{z^3+1}{4z^3 } \\[4mm]
\disp \dot{z} =     2  y \frac{z^3-1}{z^2}  \\
y(t_0) = \overline{y} \; ; \quad
z(t_0) =  \ze \; \overline{z} \;.
\end{array}
\right. \label{sis_zy}
\end{align}
Here, $ \overline{y} >0 $ and $ \overline{z} >0 $.
It is convenient to use the normalized variables $\widetilde{y}(t) =  y \left( \frac{t}{\ze^3}\right)$ and $\widetilde{z} (t) = \frac 1\ze z \left( \frac{t}{\ze^3}\right)$. 
With this transformation the small parameter $\ze$ is removed from the initial condition and  appears explicitly in the evolution equations. 
 We obtain
 \begin{align}
 \left\{
 \begin{array}{l}
 \dot{\widetilde{y}} =\disp  -  \frac{ 1}{4  {\widetilde{z}}^3 } \left(   \frac{\ze^3{\widetilde{z}}^3-1 }{3 \zh_0^2}\right)^{2/3}  +  \ze^3\left[ {\widetilde{y}}^2 + \zo^2     -      \frac{ 1 }{4  }\left(\frac{{\ze^3}{\widetilde{z}}^3-1}{3 \zh_0^2}\right)^{2/3} \right]   \\[4mm]
 \disp \dot{\widetilde{z}} =  2  \widetilde{y} \left(  \ze^3 {\widetilde{z}}  -    \frac{1}{ {\widetilde{z}}^2 }      \right) \\
 \widetilde{y}(t_0) = \overline{y} \; ; \quad
\widetilde{z}(t_0) =    \overline{z} \;.
 \end{array}
 \right.\label{z-y_sys}
 \end{align} 
We write the solution by using the Hilbert expansion
 \begin{align}
  \widetilde{y} = &\widetilde{y}_0 + \ze^3    \widetilde{y}_3 + o(\ze^3) \label{Hil_expy} \\
   \widetilde{z}  = &\widetilde{z}_0 + \ze^3   \widetilde{z}_3 +o(\ze^3) \;.\label{Hil_expz}
 \end{align} 
At the leading order in $\ze$ the system \eqref{z-y_sys} simplifies to 
\begin{align*}
\left\{
\begin{array}{l}
\disp 	\dot{\widetilde{y}_0} =     -   \frac{1}{4{\widetilde{z}_0}^3\left( {3 \zh_0^2}\right)^{2/3}  } \\[4mm]
\disp 	\dot{\widetilde{z}_0} =     - 2  \frac{{\widetilde{y}_0}}{{\widetilde{z}_0}^2}
	 \\
	 \widetilde{y}(t_0) = \overline{y} \; ; \quad
	 \widetilde{z}(t_0) =    \overline{z} \;.
	\end{array}
	\right.
\end{align*}
The solution of this system can be expressed in closed form. We have 
\begin{align}
\widetilde{z}_0(t) = &  \overline{z} e^{ 4 (\widetilde{y}_0^2(t)-\overline{y}^2)  \left(3\zh_0^2\right)^{2/3}} \geq   \overline{z} e^{ -4 \overline{y}^2  \left(3\zh_0^2\right)^{2/3}}\equiv \overline{z}_m > 0\;.  \label{z_sol_anal}
\end{align}
Equation \eqref{z_sol_anal} shows that the zeroth order  trajectory $(\widetilde{y}_0(t),\widetilde{z}_0(t))$ does not  intersects the axis $\widetilde{z}_0=0$. We verify that this statement remains valid if we include also the first correction to the solution obtained by the expansion given in Eq. \eqref{Hil_expy}-\eqref{Hil_expz}. At first we derive a preliminary bound. We are interested on the behaviour of the solution close to the singularity. We focus on the part of the  zeroth order  trajectory that starts from the point $(\overline{y},\overline{z})$, reaches the axis $\widetilde{y}_0=0$ and, due to the symmetry of the equation, ends at the point $(-\overline{y},\overline{z})$. The time at which the trajectory reaches the axis $\widetilde{y}_0=0$ is
\begin{align}
\Delta_t   = & \frac{ \overline{z}^3 }{ 4\left(3\zh_0^2\right)^{2/3}}\int_0^{\overline{y}} e^{ 12 (\widetilde{y}_0^2-\overline{y}^2)  \left(3\zh_0^2\right)^{2/3}}\dif y \leq   \frac{ \overline{z}^3 \overline{y}}{ 4\left(3\zh_0^2\right)^{2/3}} \;.\label{stima_t_point}
\end{align}
We have used the bounds $|\widetilde{y}_0|\leq \overline{y}$ and $ \overline{z}_m \leq|\widetilde{z}_0|\leq \overline{z}$. The first order correction to the solution is obtained by the system
\begin{align}
\left\{
\begin{array}{l}
\disp  \dot{ \widetilde{y}_3} =     \widetilde{y}_0^2  + \zo^2 -\frac{ 1 }{12\left(3 \zh_0^2\right)^{2/3}}   \\
\disp   \dot{ \widetilde{z}_3} = 2 \left(     \widetilde{z}_0   \widetilde{y}_0   -     \frac{ \widetilde{y}_3}{\widetilde{z}_0^{2}} \right)\\
\widetilde{y}_3 (t_0) =0\; ; \quad
\widetilde{z}_3 (t_0)= 0 \;.
\end{array}
\right. \label{sis_zy2}
\end{align}
By using  Eq. \eqref{sis_zy2} we obtain that the solution of Eq.  \eqref{sis_zy} in the proximity of  the singularity has the following expansion 
 \begin{align*}
 {z}  \geq& \ze \left[\overline{z}_m - \ze^3  \left(2 \Delta_t  \overline{z} \overline{y} +  \frac{\Delta_t^2 }{\overline{z}_m^2}   \left( \overline{y}^2  + \zo^2 -\frac{ 1 }{12\left(3 \zh_0^2\right)^{2/3}}\right)  \right)\right]   +o(\ze^4) \;.
 \end{align*} 
This proves that we can find $\ze$ small enough to ensure that the trajectory does not intersect
 the axis $z=0$.


\section{Numerics}
\begin{figure}[!h]
	\begin{center}
		A)\includegraphics[width=0.8\columnwidth]{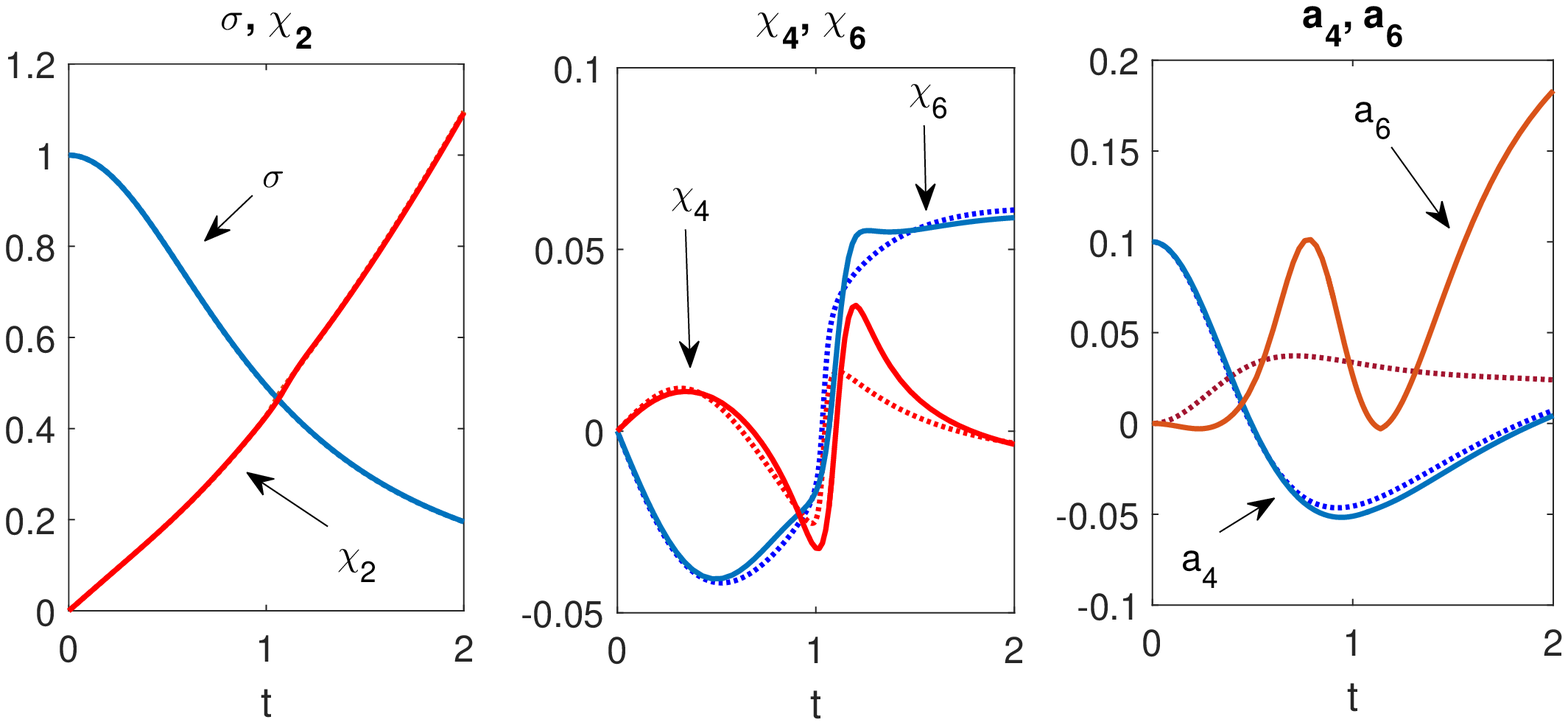}
		B)\includegraphics[width=0.8\columnwidth]{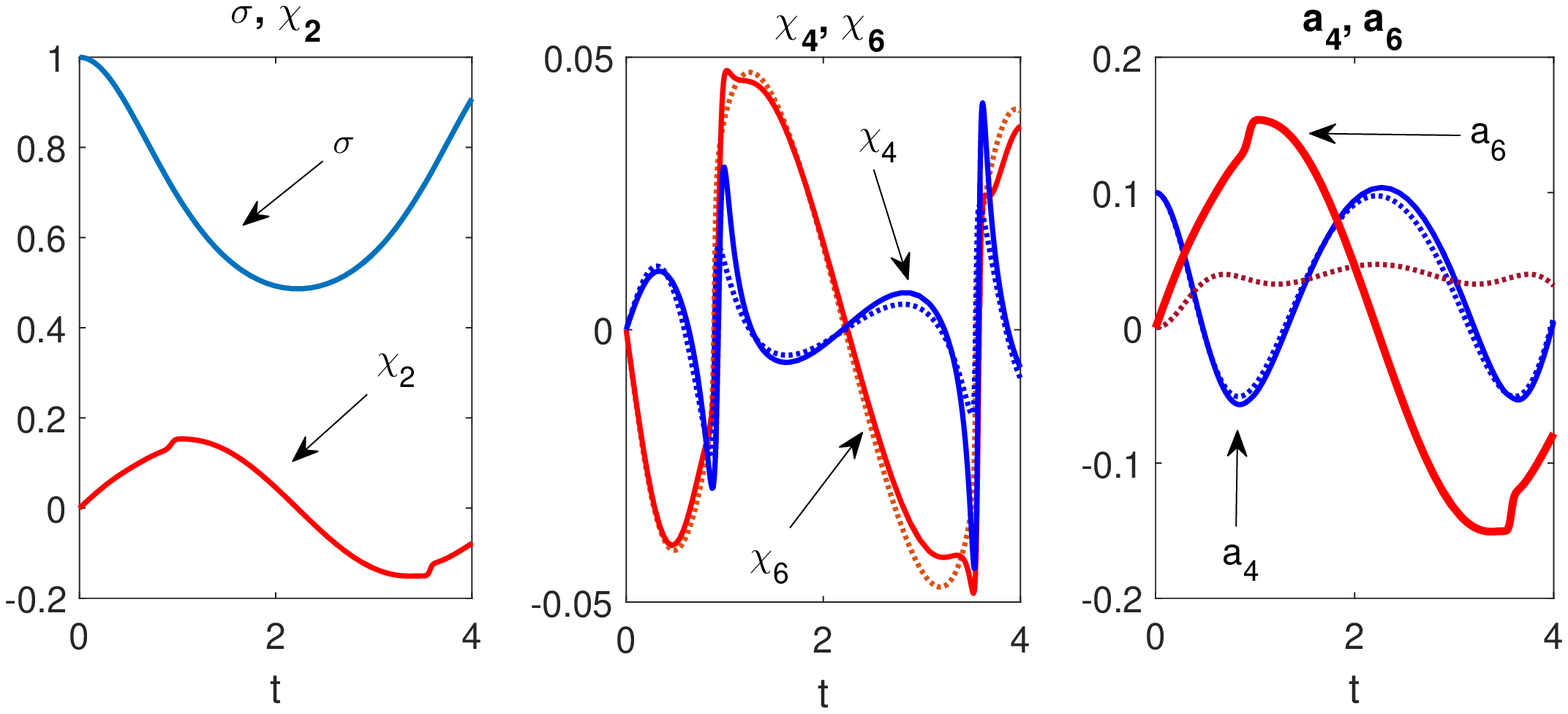}
		C)\includegraphics[width=0.4\columnwidth]{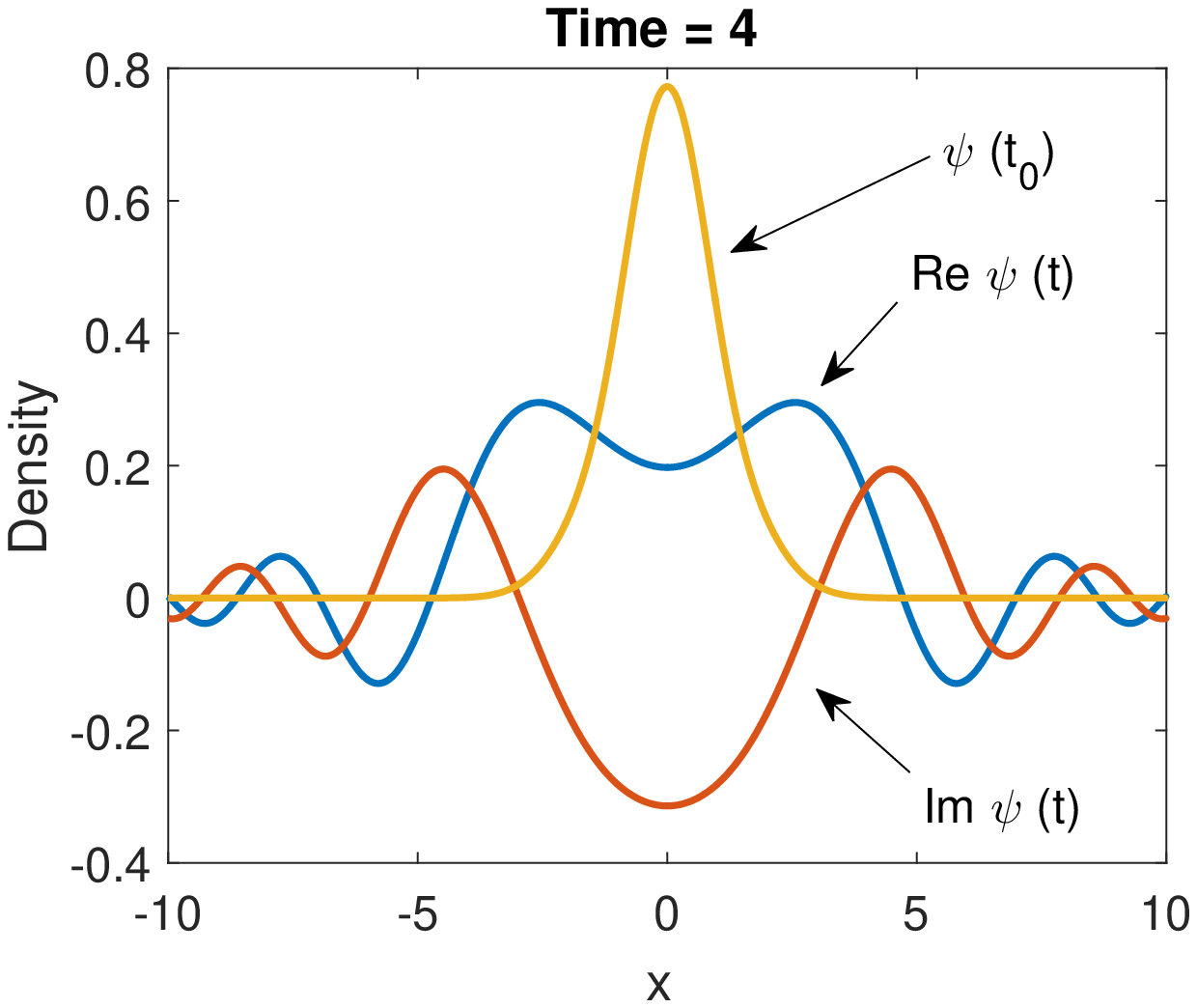}
		D)\includegraphics[width=0.4\columnwidth]{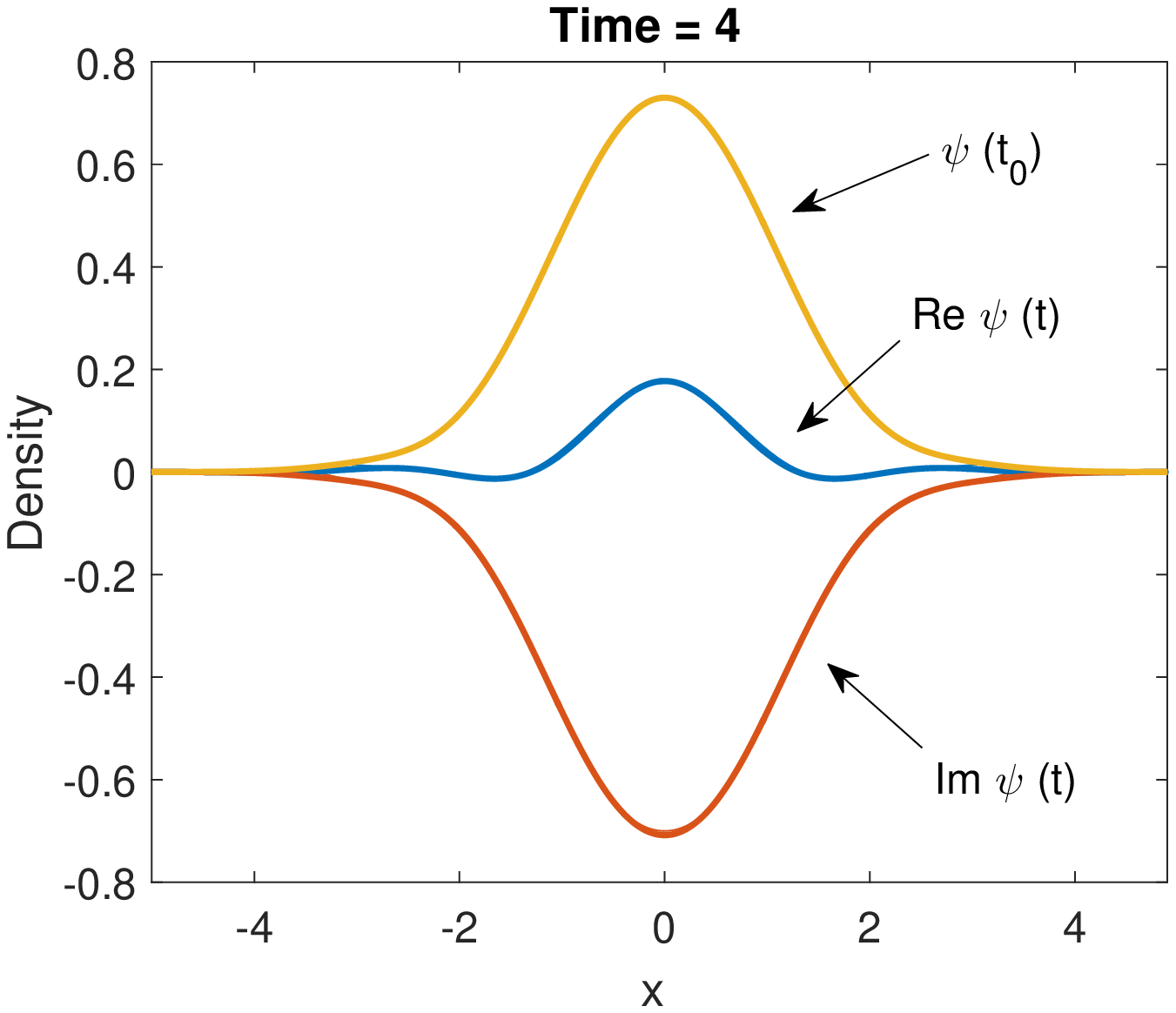}
		\caption{Numerical solution of the system \eqref{chi eq}-\eqref{s eq} compared with the solution obtained by solving the Schr\"odigner equation (dotted line). Panels A: free evolution case. Panels C: Harmonic oscillator with $\zo=0.5$. Panel C (D): Representation of the solution by the wave function for $t=t_0+4$ for the free evolution (harmonic oscillator) case. \label{fig_comp_sch} }
	\end{center}
\end{figure}

We validate the final system of equations \eqref{chi eq}-\eqref{s eq} by performing numerical tests. 
In Fig. \ref{fig_comp_sch} we depict the numerical solution of the solution (solid lines) and we compare with the solution obtained by solving directly the Schr\"odigner equation \eqref{sch ini} (dotted line). The initial condition is $\zs(t_0)=1$, $a_3(t_0)=0.1$ and the other coefficients are set to zero. In the panels A we consider the free evolution of the quantum particle and in the panels B the evolution in the presence of the harmonic potential $U(x)=\frac{\zo_0^2x^2}{2}$ with $\zo_0^2=0.5$. We find good agreement with the  direct solution of the Schr\"odinger equation especially for parameters $a_n $ and $\chi_n$ with small $n$ (we found practically no difference for the coefficients $\zs$ and $\chi_2$). Finally, in the panels C and D we depict the particle wave function at the final time $t=t_0+4$ respectively for the free evolution and the harmonic oscillator case.

\section{Bohm dynamics}\label{sec Bohm pot}
We discuss the connection of our procedure with the quantum hydrodynamic formalism based on the definition of the so called Bohm potential. In particular, we show how the Bohm potential can be expressed by our set of variables. The Bohm mechanics can be viewed as the physical interpretation of the quantum hydrodynamic equations which are  obtained by applying the Madelung transformation at the Schr\"odinger equation. The Madelung transformation consists of representing the particle wave function in polar coordinates 
\begin{align*}
\zy (x) = \sqrt{n (x-s) } e^{ i \chi (x-s)}\;.
\end{align*}
In order agree with our previous notations, we have maintained the translation of amount $s$ in the spatial coordinate. The function $\zy$ is the single particle wave function and $n(x)$ represents the particle density. According to our notations $n$ is written as
\begin{align*}
n (x) =P (x)  e^{ - x^2 \zs}\;.
\end{align*}
In this section, for the sake of simplicity we will assume $s=0$. The Madelung transformation leads to the well known quantum hydrodynamic equations 
\begin{align}
\dpp{n }{t} &=  -  \dpt{J}{x}\\
\dpp{J}{t} &=  - \dpt{ }{x} \left(\frac{J^2}{n}\right) + n \dpt{ }{x} \left(Q+U\right)\;.\label{J Mad pure state}
\end{align}
Here, $ J  =  n \dpt{\chi}{x}$ is the quantum current  and $Q(x)$ is the  quantum Bohm potential
\begin{align}
Q\equiv  &  \frac{1}{2} \frac{\dst{\sqrt{n}}{x}   }{\sqrt{n}}=  \frac{1}{4n }\dst{ n}{x} -  \frac{1}{8}     \frac{1}{n^{2}}\left(\dpt{ n}{x}   \right)^2\nn \\
=&\frac{1}{4}\left[ \frac{1}{P}\dst{P}{x} - \frac{1}{2P^2}\left(\dpt{ P}{x}   \right)^2 -2x\zs \frac{1}{P}\dpt{P}{x}+2\zs(x^2 \zs -1)\right]\;.
\label{def Q pot}
\end{align}
The expectation value of the Bohm potential is 
\begin{align*}
\int_\mathbb{R}Q (x) |\zy|^2\dif x = &\int_\mathbb{R}\left(  \frac{1}{4} \dst{P}{x} - \frac{x\zs }{2} \dpt{P}{x}+ \frac{\zs(x^2 \zs -1)P}{2}   \right)e^{ -x^2 \zs}\dif x -  \frac{1}{8}\int_\mathbb{R}\frac{1}{P}\left(\dpt{ P}{x}   \right)^2 e^{ -x^2 \zs}\dif x\;.
\end{align*}
The last term of the equation  contains the non regular part of the Bohm potential (see Eq. \eqref{bohm_pot_1}).  In the intervals where the density $P(x)$ goes to zero this term is source of numerical troubles. From  Eq. \eqref{J Mad pure state} with some algebra we obtain to the evolution equation for the phase $\chi$  
\begin{align}
\dpp{\chi}{t} &= -   \frac{1}{2} \left(\dpt{\chi}{x}   \right)^2  +  Q+U\;.\label{chi evol Madelung}
\end{align}
As a possible alternative to our variational procedure,  Eq. \eqref{chi evol Madelung} can be taken as a starting point to derive the evolution equation of the parameters $\chi_n$  obtained by expanding the function $\chi(x)$ on the Hermite polynomials.  
We multiply Eq. \eqref{chi evol Madelung} by $h_n^{\zs} e^{- x^2 \zs}$ and we integrate. We obtain 
\begin{align*}
\dpp{\chi_n}{t}= &-   \pi^{-1/4}\frac{1}{2} \int_\mathbb{R} \left(\dpt{\chi}{x}   \right)^2  h_n^{\zs} e^{- x^2 \zs}\dif x+\pi^{-1/4}
\int_\mathbb{R}\left(  Q+U\right)h_n^{\zs} e^{- x^2 \zs}\dif x\\
&-\dpt{\zs}{t}
\sum_{r}\chi_r\int_\mathbb{R}h_n^{\zs}\dpt{h_r^{\zs}}{\zs}   e^{- x^2 \zs}\dif x\;.
\end{align*}
For the first term we have (see Eq. \eqref{int p d chi})  
\begin{align*}
-  \frac{1}{2 \pi^{1/4}} \int_\mathbb{R} \left(\dpt{\chi}{x}   \right)^2  h_n^{\zs} e^{- x^2 \zs}\dif x&=-   \zs^{5/4} \sum_{r,s}\sqrt{ r s} \chi_r \chi_s      A_{n,r-1,s-1}\;.
\end{align*}
For the other terms we proceed as in Eq. \eqref{bohm_pot_1}. At first, we note that 
\begin{align*}
 \frac{1}{P}\dst{P}{x} - \frac{1}{2P^2}\left(\dpt{ P}{x}   \right)^2 \ =  \dst{R}{x}+\frac{1}{2}  \left(\dpt{R}{x} \right)^2  \;.
\end{align*}
We have introduced the auxiliary variable $R(x)=\ln(P(x))$. The integral containing the Bohm potential becomes
\begin{align*}
\int_\mathbb{R}Q h_n^{\zs} e^{- x^2 \zs}\dif x= &\int_\mathbb{R}\left(  \frac{1}{4} \dst{R}{x}+ \frac{1}{8} \left(\dpt{R}{x} \right)^2   -\frac{x\zs}{2}  \dpt{R}{x}+\frac{\zs}{2}(x^2 \zs -1)\right)h_n^{\zs} e^{- x^2 \zs}\dif x
\;.
\end{align*}
By using the expansion $R(x) = \pi^{1/4} \zs^{-1/4}\sum_m R_m h_m^\zs (x)$, we obtain the explicit form of the previous terms. For the sake of completeness, we give the result of the computations 
\begin{align*}
\frac{1}{4\pi^{ 1/4} }\int_\mathbb{R}  \dst{R}{x} h_n^{\zs} e^{- x^2 \zs}\dif x 
=&\frac{\zs^{3/4} }{2}  \sum_m R_m \sqrt{m(m-1)}\int_\mathbb{R}  h_{m-2}^\zs h_n^{\zs} e^{- x^2 \zs}\dif x 
\\
=&\frac{\zs^{3/4}  }{2}   R_{n+2} \sqrt{(n+2)(n+1)} \\
  \frac{1}{8\pi^{ 1/4}}\int_\mathbb{R}   \left(\dpt{R}{x} \right)^2   h_n^{\zs} e^{- x^2 \zs}\dif x= &  \frac{ \zs^{3/4} }{4}   \sum_{r,s}\sqrt{ r s} R_r R_s      A_{n,r-1,s-1} \\
 -\frac{1}{2\pi^{ 1/4}} \int_\mathbb{R} x\zs  \dpt{R}{x} h_n^{\zs} e^{- x^2 \zs}\dif x
= & -\frac{ \zs^{3/4}  }{2}     \left(  \sqrt{(n+1)(n+2)   } R_{n+2}+ n R_n    \right) \\
 \frac{  \zs}{2\pi^{ 1/4}}\int_\mathbb{R}(x^2 \zs -1) h_n^{\zs} e^{- x^2 \zs}\dif x
=&\frac{ \zs^{3/4}}{4} \left( \sqrt{ 2}  \zd_{n,2} - \zd_{n,0} \right)\\
\sum_{r}\chi_r\int_\mathbb{R}h_n^{\zs}\dpt{h_r^{\zs}}{\zs}   e^{- x^2 \zs}\dif x=&
\frac{2n+1}{4\zs}\chi_n  +\frac{\sqrt{(n+2) (n+1) }}{2\zs}  \chi_{n+2}^\zs   \;.
\end{align*}
%
%
%
In conclusion
  \begin{align*}
  \dpt{\chi_n}{t} &=  \sum_{r,s} \sqrt{ r s} A_{n,r-1,s-1}\left(-   \zs^{5/4} \chi_r \chi_s     + \frac{\zs^{3/4}}{4}       R_r R_s   \right)+\pi^{-1/4}
  \int_\mathbb{R}U h_n^{\zs} e^{- x^2 \zs}\dif x\\
  & -   \frac{\zs^{3/4}}{2}  n R_n    +\frac{1}{4} \zs^{3/4}\left( \sqrt{ 2}  \zd_{n,2} - \zd_{n,0} \right) -\dpt{\zs}{t}\left(  \frac{2n+1}{4\zs}\chi_n  +\frac{\sqrt{(n+2) (n+1) }}{2\zs}  \chi_{n+2}^\zs  \right)
  \end{align*}
  which agrees with Eq. \eqref{chi eq}. The evolution equation for $a_n$ and $\zs$ can be obtained by using the continuity equation. It is easy to verify that 
 \begin{align*}
 \dpp{n }{t} &=  -  \dpt{J}{x} =-  \dpt{n}{x}  \dpt{\chi}{x} -  n\dst{\chi}{x}  
 \end{align*}
 leads to
  \begin{align*}
 \left(\dpt{P}{t} -x^2 P  \dpt{\zs}{t} \right) e^{-x^2 \zs} &=-  \dpt{}{x}\left(P  \dpt{\chi}{x}   e^{-x^2 \zs} \right) \;.
  \end{align*}
 We proceed by considering the expansion on Hermite polynomial of $P$. The equation
\begin{align*}
   \int_\mathbb{R}h_n^{\zs} (x)\left(\dpt{P}{t} -x^2 P  \dpt{\zs}{t} \right) e^{-x^2 \zs}\dif x &= \int_\mathbb{R}\dpt{h_n^{\zs}}{x}     P  \dpt{\chi}{x}   e^{-x^2 \zs}  \dif x 
\end{align*}
gives
\begin{align}
 &  \sum_{r=0}^\infty\int_\mathbb{R}h_n^{\zs} (x)\left(a_r  \dpt{\zs }{t} \dpp{ h_r^\zs (x) }{\zs}  + h_r^\zs (x)  \dpt{a_r }{t}\right) e^{-x^2 \zs}\dif x -  \dpt{\zs}{t} \sum_{r=0}^\infty a_r\int_\mathbb{R}h_n^{\zs} (x)   x^2 h_r^\zs (x)    e^{-x^2 \zs}\dif x \nn \\&=\pi^{1/4}  \sum_{r,s=0}^\infty  \chi_s a_r\int_\mathbb{R}\dpt{h_n^{\zs}}{x}    h_r^\zs (x)    \dpt{ h_s^\zs (x)}{x}  e^{-x^2 \zs}  \dif x \;. \label{cal_dir_eq_a}
 \end{align}
 We have
 \begin{align*}
   \sum_{r=0}^\infty\int_\mathbb{R}h_n^{\zs} (x)\left(a_r  \dpt{\zs }{t} \dpp{ h_r^\zs (x) }{\zs} \right) e^{-x^2 \zs}\dif x  
  &=   \dpt{\zs }{t}  \left(  \frac{2n+1}{4\zs}  a_n  +\frac{\sqrt{(n+2) (n+1) }}{2\zs}  a_{n+2}   \right)  \;.
  \end{align*}
Moreover, 
\begin{align*}
     \sum_{r=0}^\infty a_r\int_\mathbb{R}h_n^{\zs} (x)   x^2 h_r^\zs (x)    e^{-x^2 \zs}\dif x =&     \frac{1}{\zs}   \sum_{r=0}^\infty a_r\int_\mathbb{R}\left(x\sqrt{\zs} h_n^{\zs} (x) \right)    \left(x\sqrt{\zs} h_r^\zs (x) \right)   e^{-x^2 \zs}\dif x  \\ 
       = &  \frac{1}{2\zs} \left[ a_n ( 2n+1 )   +a_{n-2} \sqrt{n(n-1)}           +a_{n+2}   \sqrt{(n+1)(n+2) }         \right]   \;.
\end{align*}
The last term of Eq. \eqref{cal_dir_eq_a} gives  
  \begin{align*}
 \sum_{r,s=0}^\infty  \chi_s a_r\int_\mathbb{R}\dpt{h_n^{\zs}}{x}    h_r^\zs (x)    \dpt{ h_s^\zs (x)}{x}  e^{-x^2 \zs}  \dif x  
 &= 2\zs^{5/4}\sqrt{ n }\sum_{r,s=0}^\infty \sqrt{ s } \chi_s a_r\int_\mathbb{R} h_{n-1}^1   h_r^1    h_{s-1}^1  e^{-x^2}  \dif x
 \\ &= \frac{2\zs^{5/4}}{\pi^{1/4}} \sqrt{ n   }\sum_{r,s=0}^\infty \sqrt{    s } \chi_s a_r A_{n-1,r,s-1}\;.
 \end{align*}
 We obtain
    \begin{align*}
    \dpt{a_n }{t}  &  = \frac{1}{2\zs} \dpt{\zs }{t}  \left(  \frac{2n+1}{2 }  a_n  +a_{n-2} \sqrt{n(n-1)}    \right)  + 2\zs^{5/4} \sqrt{ n   }\sum_{r,s=0}^\infty \sqrt{    s } \chi_s a_r A_{n-1,r,s-1}
 \end{align*}
 which agrees with Eq.  \eqref{an eq} for $s=0$. 
 \section{Conclusions}
 
We have presented a model designed to describe the motion of nearly localized particles. From a mathematical point, such particles are characterized by wave functions localized around a Gaussian. The oscillations of the particle wave functions around the mean particle position are reproduced by polynomials. The particles motion is described by a set of time dependent parameters whose evolution equation is obtained by the Euler-Lagrange variational approach. We have discussed the analogy of our method with the Bohm approach. We have applied our method to investigate the divergences induced by the Bohm potential. In particular, in a simple case we were able to describe the evolution of the variance of the Gaussian beam by showing the existence of two classes of trajectories separated by a singularity. Finally, we have validated our method by numerical tests. 
 
 \appendix
 \section{Derivation of  Eq. \eqref{simp der R_an}}\label{sec append Rnan}
 We derive Eq. \eqref{simp der R_an}. We start with Eq.  \eqref{bohm pot}. 
 \begin{align*}
 \mathcal{B}  \equiv &    - \frac{1}{8}\int_\mathbb{R} \frac{1}{P} \left( \dpt{P}{x} \right)^2     e^{- x^2 \zs  }  \dif x =    -  \frac{\zs^{3/4}}{4}   \sum_{ n =1 }^\infty   n \;  a_n R_n\;.
 \end{align*}
 We derive with respect to $a_n$ and we use the expansion  \eqref{expans P}
 %
 %
 \begin{align*}
    \dpt{ \mathcal{B}}{a_n}  
 =&    - \frac{1}{4 \pi^{1/4}}\int_\mathbb{R}  \dpt{R}{x} \dpt{h_n^\zs}{x}      e^{- x^2 \zs  }  \dif x +   \frac{1}{8\pi^{1/4}} \int_\mathbb{R}  \left( \dpt{R}{x}   \right)^2     h_n^\zs     e^{- x^2 \zs  }  \dif x \;,
 \end{align*}
 where we used (remind that $R=\ln(P)$)
 \begin{align*}
 \dpt{}{a_n}\left[   P(x)\left(  \dpt{  R}{x} \right)^2\right] =&  2\dpt{  R}{x} \dpt{ P(x)}{a_n}  - \dpt{  P(x)}{a_n} \left(  \dpt{  R}{x} \right)^2\;.
 \end{align*}
 Proceeding as in Eq. \eqref{int p d chi} 
 \begin{align*}
 \frac{1}{8\pi^{1/4}} \int_\mathbb{R}  \left( \dpt{R}{x}   \right)^2     h_n^\zs     e^{- x^2 \zs  }  \dif x = &  
 \frac{ \pi^{ 1/4}}{8\zs^{1/2}}  \sum_{r,s}R_r R_s \int_\mathbb{R}    \dpt{h_r^\zs (x) }{x}  \dpt{h_s^\zs (x) }{x}       h_n^\zs     e^{- x^2 \zs  }  \dif x \\
 = &  
 \frac{\pi^{ 1/4} \zs^{1/2} }{4}\sum_{r,s}  \sqrt{ r  s} R_r R_s \int_\mathbb{R}      h_{r-1}^\zs   h_{s-1}^\zs      h_n^\zs     e^{- x^2 \zs  }  \dif x \\
 = &  
 \frac{\zs^{3/4}}{4}  \sum_{r,s}  \sqrt{ r  s} R_r R_s  \pi^{ 1/4}\int_\mathbb{R}      h_{r-1}^1 h_{s-1}^1      h_n^1     e^{- x^2  }  \dif x \\
 =& \frac{\zs^{3/4}}{4} \sum_{r,s\geq 1} \sqrt{rs} R_rR_s A_{n,r-1,s-1} \;,
 \end{align*}
 and (we use Eq. \eqref{der hn-zs})
 \begin{align*}
   \frac{1}{4 \pi^{1/4}}\int_\mathbb{R}  \dpt{R}{x} \dpt{h_n^\zs}{x}      e^{- x^2 \zs  }  \dif x =&     \frac{\zs^{3/4}}{2}  \sum_s   \sqrt{ sn }R_s \int_\mathbb{R} h_{s-1}^\zs     h_{n-1}^\zs     e^{- x^2 \zs  }  \dif x =  \frac{\zs^{3/4}}{2}  nR_n \;.
 \end{align*}
 Finally
 \begin{align*}
  \dpt{\mathcal{B}}{a_n}   =& -  \frac{ \zs^{3/4}}{2}  nR_n+ \frac{\zs^{3/4}}{4} \sum_{r,s\geq 1} \sqrt{rs} R_rR_s A_{n,r-1,s-1} \;.
 \end{align*}

\end{document}